\documentclass[review]{elsarticle}

\usepackage{lineno,hyperref}
\usepackage{amsmath}
\usepackage{amssymb}
\usepackage{cancel}
\usepackage{bm}
\usepackage{caption}
\usepackage{subcaption,color}
\usepackage{xcolor}
\usepackage{soul}
\usepackage{verbatim}  
\usepackage{makecell}
\usepackage{adjustbox}
\usepackage[noabbrev]{cleveref}
\biboptions{sort&compress}
\modulolinenumbers[5]

\journal{Journal of Computational Physics}

\newcommand{\sigmat}{\sigma_\mathrm{t}}
\newcommand{\sigmas}{\sigma_\mathrm{s}}
\newcommand{\sigmaf}{\sigma_\mathrm{f}}

\newcommand{\PN}{P$_N$}
\renewcommand{\vec}[1]{\bm{#1}} 
\newcommand{\vd}{\bm{\cdot}} 
\newcommand{\grad}{\vec{\nabla}} 

\begin{document}

\begin{frontmatter}

\title{A low-rank method for two-dimensional time-dependent radiation transport calculations} 

\author[mymainaddress]{Zhuogang Peng}
\ead{zpeng5@nd.edu}

\author[mymainaddress]{Ryan G.\ McClarren\corref{mycorrespondingauthor}}
\cortext[mycorrespondingauthor]{Corresponding author}
\ead{rmcclarr@nd.edu}
 
\author[mysecondaryaddress]{Martin Frank}
\ead{martin.frank@kit.edu}

\address[mymainaddress]{Department of Aerospace and Mechanical Engineering, University of Notre Dame, Notre Dame, IN, 46545, USA}
\address[mysecondaryaddress]{Karlsruhe Institute of Technology, Steinbuch Centre for Computing, Karlsruhe, Germany}

\begin{abstract}
The low-rank approximation is a complexity reduction technique to approximate a tensor or a matrix with a reduced rank, which has been applied to the simulation of high dimensional problems to reduce the memory required and computational cost. In this work, a dynamical low-rank approximation method is developed for the time-dependent radiation transport equation in 1-D and 2-D Cartesian geometries. Using a finite volume discretization in space and a spherical harmonics basis in angle, we construct a system that evolves on a low-rank manifold via an operator splitting approach. Numerical results on five test problems demonstrate that the low-rank solution requires less memory than solving the full rank equations with the same accuracy. It is furthermore shown that the low-rank algorithm can obtain high-fidelity results at a moderate extra cost by increasing the number of basis functions while keeping the rank fixed.
\end{abstract}

\begin{keyword}
Low-rank approximation, Radiation transport
\end{keyword}

\end{frontmatter}


\section{Introduction}
The numerical simulation of the radiation transport process is fundamental in a wide range of applications from supernovas to medical imaging, where accurate numerical solutions of a linear Boltzmann equation, also known as the radiative transfer equation (RTE), are required. In this equation, we are solving for the specific intensity, which is a seven-dimensional function that describes the movement of the flow of particles in terms of time, position, direction, and energy. Despite the rapid development of high-performance computing, the numerical solution of the RTE remains challenging due to the high computational costs and the large memory requirements caused by such a high-dimensional phase space. In this paper, we focus on reducing the memory required to solve the RTE to enable better performance on exascale-class computers \cite{Shalf2011}.

Due to the rich phase space, computational methods for radiation transport require discretizations in energy and direction\footnote{The terminology direction and angle are typically used interchangeably to describe the angular component of the phase space.}, in addition to spatial and temporal discretizations.  For the energy variable, the multigroup method is commonly used \cite{Pomraning}. The direction, or angular variables, can be treated by solving the equations along particular directions via the discrete ordinates or S$_N$ method \cite{LewisMiller}. An alternative technique employs a basis expansion using the natural basis for the sphere: spherical harmonics. The spherical harmonics or P$_N$ method uses a truncated expansion to approximate the angular dependence \cite{Mcclarren2008, Laboure:2016hm, Dargaville:2019fa}. Low-order approximations are also used such as flux-limited diffusion \cite{morel2000, Olson:2000vq}, simplified P$_N$ \cite{McClarren:2011ga, Frank:2007ig,Modest:2012df} along with hybrid methods such as quasi-diffusion \cite{anistratov1986solution}. Finally, stochastic methods based on the implicit Monte Carlo method \cite{Fleck:1971vj,McClarren:2009bs, Wollaber:2016hv,SmedleyStevenson:2015gt} are an alternative approach, though often requiring significant computational cost.

Although the aforementioned methods require many degrees of freedom to describe the phase-space dependence of the solution, it is known that many transport problems require only a subspace of the full phase space (called a manifold in mathematical parlance) to describe the transport of particles. An example of this phenomenon is problems in the diffusion limit: these problems require only a linear dependence on angular variables.  One can also formulate problems where this manifold over which the solution depends evolves over time: a beam entering a scattering medium would be described by a delta-function in space and angle at time zero, but eventually, relax to much smoother distribution that we could characterize using a simple basis expansion.

We desire to generalize this idea, and possibly automatically discover the manifold that describes the system evolution. We accomplish this task by expressing the solution to a transport problem as a basis expansion in space and angle and using techniques to determine what subspace of those bases are needed to describe the solution and how that subspace evolves. We use the dynamical low-rank approximation (DLRA) of Koch and Lubich to evolve time-dependent matrices by tangent-space projection \cite{Koch2007}. DLRA has been extended to tensors \cite{Koch2010}, and further results can be found in \cite{Nonnenmacher2008}. DLRA has been used to reduce the computational complexity of quantum propagation \cite{Kloss2017} by restricting the evolution to lower-rank. 

{Our work builds on this and other previous work to demonstrate how low-rank methods can improve radiative calculations. An earlier paper applied the low-rank approximation to solve the Vlasov equation {\cite{KatharinaKormann2015}}. However, in that case, the rank is not fixed, so the singular value decomposition is required in each time step to truncate the solution to the designated rank. In this work, we apply the dynamical low-rank projector splitting technique developed by Lubich and Oseledets {\cite{Lubich2014}} to neutral particle transport. A similar algorithm has been applied to plasma dynamics {\cite{Einkemmer2018}\cite{Einkemmer2019}\cite{Einkemmer2019_vlasovMaxwell}} and computational fluid dynamics {\cite{Einkemmer2019_weaklyflow}}. }

{In terms of low-rank methods for radiative transfer calculations, we presented preliminary results using DLRA at a recent conference {\cite{peng2019low}}. Additionally, the asymptotic analysis of this splitting method in a one-dimensional radiative transfer equation with the backward Euler and Crank-Nicolson scheme was made in {\cite{ding2019error}}. Our work presents a detailed numerical scheme for 1-D and 2-D problems and we specifically study the required storage during the solution process.}

\subsection{Overview of Projector-Splitting Method}
Here we give a brief mathematical introduction to the robust and accurate projector-splitting method {{\cite{Lubich2014}}} to perform the DLRA for matrix differential equations of the form
\begin{linenomath*}\[
\frac{\partial}{\partial t} A(t)  \equiv \dot{A}(t)= F(A(t)),
\]\end{linenomath*}
for $A(t) \in \mathbb{R}^{m \times n}$.
DLRA seeks to find an approximating matrix $Y(t)=U(t)S(t)V^T(t)$, where $U(t)\in\mathbb{R}^{m\times r}$ , $V(t)\in\mathbb{R}^{n\times r}$ are orthonormal matrices and $S(t)\in\mathbb{R}^{r\times r}$ so that $Y(t)$ has rank $r$. Note that rank $r$ matrices are a manifold, $\mathcal{M}_r$, in the space $\mathbb{R}^{m \times n}$. We evolve the solution on $\mathcal{M}_r$ directly by inserting the ansatz for $Y(t)$ into the equation and projecting the right-hand side onto the tangent space of $\mathcal{M}_r$. 

In this work, we develop a reduced system of the radiative transfer equation with this low-rank method, where the intensity only involves its low-rank orthogonal bases rather the full phase space. We design a numerical scheme for the 2-D RTE with a finite volume discretization in space and a spherical harmonic (P$_N$) expansion in angle. We further demonstrate the memory saving features by comparing the low-rank results with full-rank solutions and benchmark results.

\section{Derivation of Method}
We consider {an} energy-independent radiation transport equation
\begin{linenomath*}\begin{equation} \label{RadiativeTransfer} 
\frac{1}{c} \frac{\partial \psi (\vec{r}, \hat{\Omega}, t)}{\partial t} + \hat{\Omega} \vd \grad \psi (\vec{r}, \hat{\Omega}, t)  +  \sigmat(\vec{r}) \psi (\vec{r}, \hat{\Omega}, t) 
= \frac{1}{4\pi} \sigmas(\vec{r}) \phi(\vec{r}, t) + Q(\vec{r},t).
\end{equation}\end{linenomath*}
Here the angular flux $\psi(\vec{r}, \hat{\Omega}, t)$ with units of particles per area per steradian per time is a function of position $\vec{r} \in D$ where $D$ is the computational domain, direction $\hat{\Omega}(\mu, \varphi)$ where $\mu$ is the cosine of the polar angle and $\varphi$ is the azimuthal angle, and time $t$. Note that $\hat{\Omega}$ is a unit vector. The total and isotropic scattering macroscopic cross-sections with units of inverse length are denoted as $\sigmat(\vec{r})$ and $\sigmas(\vec{r})$, respectively, $c$ is the particle speed and $Q(\vec{r},t)$ is a prescribed source with units of particles per volume per time. We also write the scalar flux, $\phi(\vec{r},t)$, as the integral of the angular flux
\begin{linenomath*}\begin{equation}
\phi(\vec{r},t) = \int_{4 \pi}\psi(\vec{r}, \hat{\Omega}, t) \, d \hat{\Omega}.
\end{equation}\end{linenomath*}

In this study we approximate the solution to Eq.~\eqref{RadiativeTransfer}  using the form
\begin{linenomath*}\begin{equation}\label{LRA1}
\psi(\vec{r},\hat{\Omega},t) \approx \sum_{i,j = 1}^r X_i(\vec{r},t)S_{ij}(t)W_j(\hat{\Omega},t);
\end{equation}\end{linenomath*}
that is, we seek the best approximation with rank $r$ to the solution of Eq.~(\ref{RadiativeTransfer}), where we have written
$X_i$ as an orthonormal basis for $\vec{r}$ and $W_j$ as an orthonormal basis for $\hat{\Omega}$ using the inner products 
\begin{linenomath*}\[\langle f,g\rangle _{\vec{r}} = \int_{D} f  g\, d\vec{r}, \qquad \langle f,g\rangle _{\hat{\Omega}} = \int_{4 \pi} f g \, d\hat{\Omega} . \]\end{linenomath*} 
Due to orthonormality we also have  $\langle X_i, {X}_j \rangle_{\vec{r}} = \langle W_i, {W}_j \rangle_{\hat{\Omega}} = \delta_{ij}.$ 
We also use $
\bar{X} = \{X_1, X_2, ..., X_r\}
$
and 
$
\bar{W} = \{W_1, W_2, ..., W_r\}
$ as ansatz spaces. 
The expansion in Eq.~\eqref{LRA1} is not unique and we add orthogonal constraints  $\langle X_i, \dot{X}_j \rangle_{\vec{r}} = 0$ and $\langle W_i, \dot{W}_j \rangle_{\hat{\Omega}}  = 0$ as gauge conditions, which identify $X$ and $W$ in Grassmann manifolds \cite{Edelman1998}.
We now define orthogonal projectors using the bases:
\begin{linenomath*}\begin{equation}\label{Projection1}
P_{\bar{X}} g = \sum_{i=1}^{r}X_i \langle X_i g \rangle_{\vec{r}},
\end{equation}\end{linenomath*}
\begin{linenomath*}\begin{equation}\label{Projection2}
P_{\bar{W}} g = \sum_{j=1}^{r}W_j \langle W_j g \rangle_{\hat{\Omega}}.
\end{equation}\end{linenomath*}
 
{The projector onto the tangent space of the low-rank manifold {$\mathcal{M}_r$} is given in {\cite{Koch2007}} as} 
\begin{linenomath*}\begin{equation}\label{projector0} {P_g = P_{\bar{W}}g  - P_{\bar{X}} P_{\bar{W}} g + P_{\bar{X}} g} .\end{equation}\end{linenomath*}
We apply the projectors to define a {Lie splitting} of the original equations into three steps; each of these is solved {consecutively}
for a time step : 
\begin{linenomath*}\begin{multline}\label{eq: MainEq1}
\partial_t {\psi}(\vec{r}, \hat{\Omega},t)  = P_{\bar{W}} \bigg( - \hat{\Omega} \cdot \grad \psi(\vec{r}, \hat{\Omega}, t)
- \sigma_t(\vec{r}, t) \psi (\vec{r}, \hat{\Omega}, t) \\
+ \frac{1}{4\pi} \sigmas(\vec{r}, t) \phi(\vec{r}, t) 
+ Q(\vec{r},t) \bigg),
\end{multline}\end{linenomath*}

\begin{linenomath*}\begin{multline}\label{eq: MainEq2}
\partial_t {\psi}(\vec{r}, \hat{\Omega},t) = - P_{\bar{X}} P_{\bar{W}} \bigg( - \hat{\Omega} \cdot \grad \psi(\vec{r}, \hat{\Omega}, t)
- \sigma_t(\vec{r}, t) \psi (\vec{r}, \hat{\Omega}, t) \\
+ \frac{1}{4\pi} \sigmas(\vec{r}, t) \phi(\vec{r}, t) 
+ Q(\vec{r},t) \bigg),
\end{multline}\end{linenomath*}

\begin{linenomath*}\begin{multline}\label{eq: MainEq3}
\partial_t {\psi}(\vec{r}, \hat{\Omega},t) = P_{\bar{X}} \bigg( - \hat{\Omega}\cdot \grad \psi(\vec{r}, \hat{\Omega}, t)
- \sigma_t(\vec{r}, t) \psi (\vec{r}, \hat{\Omega}, t) \\
+ \frac{1}{4\pi} \sigmas(\vec{r}, t) \phi(\vec{r}, t) 
+ Q(\vec{r},t) \bigg).
\end{multline}\end{linenomath*}

{Thus, equation {\eqref{eq: MainEq2}} uses the solution of equation {\eqref{eq: MainEq1}} as an initial condition, and the equation {\eqref{eq: MainEq3}} uses the solution of equation {\eqref{eq: MainEq2}} as an initial condition.} It can be shown that the above evolution is contained in the low-rank manifold $\mathcal{M}_r$, if the initial value is in $\mathcal{M}_r$ \cite{Lubich2014} because the right-hand side of each step remains in the tangent space $\mathcal{T} \mathcal{M}_r$. 
    
The main advantage of this scheme stems from the fact that only the low-rank components $X(\vec{r}, t)$, $S(t)$, and $W(\hat{\Omega}, t)$ need to be stored during the time evolution rather than the full size solution $\psi(\vec{r}, \hat{\Omega}, t)$. To demonstrate this we first formulate the projections (\ref{eq: MainEq1} - \ref{eq: MainEq3}) explicitly from time $t_0$ to $t_0+h$ where $h$ is the step size. The low-rank representation of $\psi(\vec{r},\hat{\Omega},t_0)$ is given by the initial condition
\begin{linenomath*}\begin{equation}\label{eq: ini_cond}
\psi^{(0)}(\vec{r},\hat{\Omega},t_0) = \sum_{i,j = 1}^r X^{(0)}_i(\vec{r},t_0)S^{(0)}_{ij}(t_0)W^{(0)}_j(\hat{\Omega},t_0).
\end{equation}\end{linenomath*}
In the first projection the basis $W_j$ does not change with time and is evaluated at the initial value $W^{(0)}_j(\hat{\Omega})$. We simplify the notation by writing $K_j(\vec{r},t) = \sum^{r}_{i} X_{i}(\vec{r},t)S_{ij}(t)$, then \eqref{eq: ini_cond} becomes
\begin{linenomath*}\begin{equation}\label{eq: ini_cond2}
\psi^{(0)}(\vec{r},\hat{\Omega},t_0) = \sum_{j=1}^r K^{(0)}_j(\vec{r},t_0) W^{(0)}_j(\hat{\Omega}).
\end{equation}\end{linenomath*}
We plug this solution into Eq.~\eqref{eq: MainEq1} and multiply by $W^{(0)}_\ell(\hat{\Omega})$ and integrate over $\mu$ and $\varphi$ to get
\begin{linenomath*}\begin{multline}
\label{eq:psi1}
\partial_t K_j = -\sum_{l=1}^{r} \grad K_l \, \langle \hat{\Omega} W_l W_j\rangle_{\hat{\Omega}} - \sigmat K_j 
+ \frac{1}{4\pi} \sigmas \sum_{l=1}^{r}  K_l \langle W_l\rangle_{\hat{\Omega}} \langle W_j\rangle_{\hat{\Omega}} \\+ Q \langle W_j\rangle_{\hat{\Omega}}.
\end{multline}\end{linenomath*}
Equation \eqref{eq:psi1} resembles the standard P$_N$ equations, a point we will return to later. It is a system of advection problems coupled through the streaming term. 

We can then factor $K_j^{(1)}{(\vec{r},t_0+h)}$, which is the solution of Eq.~\eqref{eq:psi1} into $X_{i}^{(1)}{(\vec{r},t_0+h)}$ and $S_{ij}^{(1)}{(t_0+h)}$ using a QR decomposition or singular value decomposition. In the second step both the $X_i$ and $W_j$ are preserved.
The initial conditions to solve Eq.~\eqref{eq: MainEq2}
are $X_{i}^{(2)}(\vec{r}) = X_{i}^{(1)}(\vec{r}, t_0+h)$, $S_{i}^{(2)}(t_0) = S_{i}^{(1)}(t_0+h)$, and $W_{i}^{(2)}(\hat{\Omega}) = W_{i}^{(0)}(\hat{\Omega})$.
Then, we can perform similar calculations on Eq.~\eqref{eq: MainEq2} to get 
\begin{linenomath*}\begin{multline}\label{eq:s_eq}
\frac{d}{dt} S_{ij} = \sum_{kl}^{r} \langle \grad X_k \, X_l \rangle_{\vec{r}} S_{kl} \langle \hat{\Omega} W_l W_j \rangle_{\hat{\Omega}} 
+\sum_{k}^{r} \langle \sigma_t X_k X_i \rangle_{\vec{r}} S_{kj} \\
- \frac{1}{4\pi} \sum_{kl}^{r} \langle \sigma_s X_k X_i \rangle_{\vec{r}}S_{kl}\langle W_l\rangle_{\hat{\Omega}} \langle W_j\rangle_{\hat{\Omega}} - \langle X_i Q \rangle_{\vec{r}} \langle W_j\rangle_{\hat{\Omega}}.
\end{multline}\end{linenomath*}
We call the solution $S_{ij}^{(2)}$. Equation \eqref{eq:s_eq} defines a set of $r^2$ ordinary differential equations. The solution is used to create an initial condition for Eq.~\eqref{eq: MainEq3}, where $X_{i}^{(3)}(\vec{r}) = X_{i}^{(1)}(\vec{r}, t_0+h)$, $S_{i}^{(3)}(t_0) = S_{i}^{(2)}(t_0+h)$, and $W_{i}^{(3)}(\hat{\Omega}, t_0) = W_{i}^{(0)}(\hat{\Omega})$. Notice that $X_i$ does not change with time in this step. 

Writing $L_i = \sum_j^r S_{ij}(t)W_{j}(\hat{\Omega}, t)$ we can multiply Eq.~\eqref{eq: MainEq3} by a spatial basis function and integrate over space to get
\begin{linenomath*}\begin{multline}\label{eq:L_eq}
\partial_t L_{i} = -\hat{\Omega} \sum_{k}^{r} \langle \grad X_k \, X_i \rangle_{\vec{r}} L_k - \sum_{k}^{r} \langle \sigma_t X_k X_i \rangle_{\vec{r}} L_{k} 
+ \frac{1}{4\pi} \sum_{k}^{r} \langle \sigma_s X_k X_i \rangle_{\vec{r}} \langle L_k \rangle_{\hat{\Omega}} \\
+ \langle Q X_i \rangle_{\vec{r}},
\end{multline}\end{linenomath*}
which evolves the solution in $\mu$ and $\varphi$ spaces. Upon factoring $L_i = S^{(3)}_{ij}(t)W_j^{(3)}(\hat{\Omega},t)$ using a QR decomposition we can write the low-rank solution as $\psi(\vec{r}, \hat{\Omega}, t_0+h) = \sum_{i,j,=1}^r X_i^{(1)}(\vec{r},t_0+h) S^{(3)}_{ij}(t_0+h)W_j^{(3)}(\hat{\Omega},t_0+h).$ {Notice that there is no need to calculate and store the full solution during the evolution.}

\section{Numerical Scheme}
In this section the procedure outlined above of solving Eqs.~\eqref{eq:psi1}, \eqref{eq:s_eq}, and \eqref{eq:L_eq} is implemented in one and two spatial dimensions with a first-order explicit time integrator \cite{Lubich2014} and a finite volume discretization in space. For the angular basis, we use a spherical harmonics expansion. In this section, we describe the numerical method to solve the 2D problem in detail.

\subsection{Discretization details} 
In the two dimensional system, we write the spatial variables as $x$ and $z$. The transport equation in this reduced geometry is 
\begin{linenomath*}\begin{multline} \label{eq: 2D_RadiativeTransfer}
\frac{1}{c} \frac{\partial \psi (x,z,\mu,\varphi, t)}{\partial t} + \mu \partial_z \psi(x, z, \mu, \varphi, t) + \sqrt{1 - \mu^2} \cos \varphi \, \partial_x \psi(x, z, \mu, \varphi, t) \\
+ \sigmat(x, z) \psi (x,z,\mu,\varphi, t) 
= \frac{1}{4\pi} \sigmas(x, z) \phi(x, z, t) + Q(x,z,t).
\end{multline}\end{linenomath*}
The projection system simplifies to
\begin{linenomath*}\begin{multline}
\label{eq: 2D_k_eq}
\partial_t K_j = -\sum_{l=1}^{r} \partial_z K_{l} \, \langle \mu W_j W_{l}\rangle_{\hat{\Omega}} -\sum_{l=1}^{r} \partial_x K_{l} \, \left\langle {\sqrt{1 - \mu^2} \cos \varphi} \, W_j W_{l}\right\rangle_{\hat{\Omega}} - \sigmat K_j \\
+ \frac{\sigmas}{2}  \sum_{l=1}^{r}  K_{l} \langle W_{l} \rangle_{\mu} \langle W_j\rangle_{\mu} + \frac{\langle W_j\rangle_{\mu}}{2} Q,
\end{multline}\end{linenomath*}

\begin{linenomath*}\begin{multline}\label{eq: 2D_s_eq}
\partial_t S_{ij} = \sum_{kl}^{r} \langle \partial_z X_k \, X_l \rangle_{\vec{r}} S_{kl} \left\langle \mu W_l W_j \right\rangle_{\hat{\Omega}} + \sum_{kl}^{r} \langle \partial_x X_k \, X_l \rangle_{\vec{r}} S_{kl} \langle {\sqrt{1 - \mu^2} \cos \varphi} \,W_l W_j \rangle_{\hat{\Omega}}\\
+\sum_{k}^{r} \langle \sigma_t X_k X_i \rangle_{\vec{r}} S_{kj} 
- \frac{1}{2} \sum_{kl}^{r} \langle \sigma_s X_k X_i \rangle_{\vec{r}} S_{kl} \langle W_l\rangle_{\hat{\Omega}} \langle W_j\rangle_{\hat{\Omega}} - \frac{1}{2}\langle X_i Q \rangle_{\vec{r}} \langle W_j\rangle_{\hat{\Omega}},
\end{multline}\end{linenomath*}

\begin{linenomath*}\begin{multline}\label{eq: 2D_L_eq}
\partial_t L_{i} = -\mu \sum_{k}^{r} \langle \partial_z X_k \, X_i \rangle_{\vec{r}} L_k - {\sqrt{1 - \mu^2} \cos \varphi} \sum_{k}^{r} \langle \partial_x X_k \, X_i \rangle_{\vec{r}} L_k -  \sum_{k}^{r} \langle \sigma_t X_k X_i \rangle_{\vec{r}} L_{k} \\
+ \frac{1}{2} \sum_{k}^{r} \langle \sigma_s X_k X_i \rangle_{\vec{r}} \langle L_k \rangle_{\hat{\Omega}} 
+ \frac{1}{2}\langle Q X_i \rangle_{\vec{r}}.
\end{multline}\end{linenomath*}

The bases we use arise from a finite volume discretization in space with a constant mesh area $\Delta x  \Delta z$ and $N_x \times N_z$ zones, and the truncated at $N_l$  spherical harmonics in angle. To make orthonormal bases we define 
\begin{linenomath*}\begin{equation}\label{eq: 2D_Discretization1}
X_i(t,x,z) = \sum_{p=1}^{N_x}\sum_{q=1}^{N_z}Z_{p q}(x,z)u_{p q i}(t)
\end{equation}\end{linenomath*}

\begin{linenomath*}\begin{equation}\label{eq: 2D_Discretization2}
W_j(t, \mu, \varphi) = \sum_{l=0}^{N_l} \, \sum_{k=0}^{l}Y_{l}^{k}(\mu, \varphi) v_{l k j}(t)
\end{equation}\end{linenomath*}
The spatial basis functions are given by $Z_{pq}(x,z) = \frac{1}{\sqrt{\Delta x \, \Delta z}}$ with $x \in [x_{p-\frac{1}{2}},x_{p+\frac{1}{2}}]$ and $z \in [z_{q-\frac{1}{2}},z_{q+\frac{1}{2}}]$, where $p$ and $q$ are the cell index. {Note that $Z_{pq}(x, z) = 0$ outside the cell $\{p, q\}$.} The spherical harmonics are defined as
\begin{linenomath*}\[Y_l^k(\mu, \varphi) = \sqrt{\frac{2l+1}{4 \pi} \, \frac{(l-k)!}{(l+k)!}} \, P_l^k(\mu) \, e^{i \, k\varphi},\]\end{linenomath*}
where $P_{l} ^k (\mu)$ is the associate Legendre polynomial.  The negative $k$ are not necessary here because of the recursion properties of the spherical harmonics \cite{Brown2005}. Here $u_{pqi}$ and $v_{lkj}$ are components of the time dependent tensor $U(t) \in \mathbb{R}^{Nx \times Nz \times r}$ and  $V(t) \in \mathbb{R}^{N_l \times (N_l+1) \times r}$. Additionally, in 2-D Cartesian geometry, the imaginary part of the spherical harmonics can be decoupled and ignored \cite{Brunner2005}. 

We also notice that the computations in Eqs.~\eqref{eq: 2D_Discretization1}, \eqref{eq: 2D_Discretization2} can be performed {by performing matrix multiplications} instead of tensor operation. This is achieved by rearranging the matrices $Z \in \mathbb{R}^{N_x \times N_z}$ and $Y \in \mathbb{R}^{\frac{1}{2} \left(N_l+1\right) \left(N_l+2\right)}$ into vectors and tensors $U$ and $V$ to matrices. 

To make the notation consistent we denote the number of degrees of freedom in space as $m = N_x N_z$ and the angular degrees of freedom as   $n = \frac{1}{2}(N_l+1)(N_l+2)$. In addition, the indices of $U$ and $V$ conform to Eqs.~\eqref{eq: 2D_Discretization1} and \eqref{eq: 2D_Discretization2}.

Consequently the new ansatz spaces $X$ and $W$ have the form
\begin{linenomath*}\[
X = UZ = \begin{bmatrix}
u_{1\,1\,1}&u_{1\,1\,2}&...&u_{1\,1\,r}\\...&...&...&...\\u_{N_x\,1\,1}&u_{N_x\,1\,2}&...&u_{N_x\,1\,r}\\u_{1\,2\,1}&u_{1\,2\,2}&...&u_{1\,2\,r}\\...&...&...&...\\u_{N_x \,N_z\,1}&u_{N_x \,N_z\,2}&...&u_{N_x \,N_z\,r}
\end{bmatrix} ^T
\begin{bmatrix}
Z_{1\,1}\\...\\Z_{N_x \, 1}\\Z_{1\,2}\\...\\Z_{N_x \, N_z}
\end{bmatrix}
\]\end{linenomath*}

\begin{linenomath*}\[
W = V Y =
\begin{bmatrix}
v_{0\,0\,1}&v_{0\,0\,2}&...&v_{0\,0\,r}\\v_{0\,1\,1}&v_{0\,1\,2}&...&v_{0\,1\,r}\\v_{1\,1\,1}&v_{1\,1\,2}&...&v_{1\,1\,r}\\v_{0 \,2\,1}&v_{0\,2\,2}&...&v_{0 \,2\,r}\\...&...&...&...\\v_{N_l \,N_l\,1}&v_{N_l \,N_l\,2}&...&v_{N_l \,N_l\,r}
\end{bmatrix} ^T
\begin{bmatrix}
Y_0^0\\Y_1^{0}\\Y_1^{1}\\Y_2^{0}\\...\\Y_{N_l}^{N_l}
\end{bmatrix}
\]\end{linenomath*}

To solve Eqs.~\eqref{eq: 2D_s_eq} and \eqref{eq: 2D_L_eq} we need to calculate spatial integration terms like $\langle \partial_z X_k \, X_l \rangle_{\vec{r}}$. Due to our use of finite volume method, the basis $X_i$ is discontinuous and piecewise constant in space. 
We apply integration by parts and obtain
\begin{linenomath*}\begin{equation}\label{eq: 2D_Term1}
\begin{aligned}
\langle \partial_z X_k \, X_l \rangle_{\vec{r}} = & \left \langle \partial_z \Big(\sum_p^{N_x} \sum_q^{N_z} Z_{pq} u_{pqk} \Big)  \sum_{p'}^{N_x} \sum_{q'}^{N_z} Z_{p'q'} u_{p'q'k} \right \rangle_{\vec{r}} \\
=& \sum_{p'}^{N_x} \sum_{q'}^{N_z} u_{p'q'l} \left ( \left \langle \partial_z (Z_{p'q'} \sum_p^{N_x} \sum_q^{N_z} Z_{pq} u_{pqk}) \right \rangle _{\vec{r}} - \left \langle \cancelto{0}{\partial_z Z_{p'q'}} \quad  \sum_{p'}^{N_x} \sum_{q'}^{N_z} Z_{p'q'} u_{pqk} \right \rangle _{\vec{r}} \right )\\
=& \sum_{p}^{N_x} \sum_{q}^{N_z} u_{pql} \left ( \frac{\{u\}_{p,q+\frac12,k} - \{u\}_{p,q-\frac12,k}}{2 \Delta z} \right)
\end{aligned}
\end{equation}\end{linenomath*}
Here, $\{u\}_{p,q+\frac12,k}$ denotes the value at the cell boundary between the spatial cells $q$ and $q+1$. We set the value at the cell boundary by the average of its left and right values, such as $\{u\}_{p,q+\frac12,k} = \frac12(u_{p,q+1,k}+u_{p,q,k})$.

For the angular terms,  e.g., $ \langle \mu W_j W_{j'} \rangle_{\hat{\Omega}}$, we can  write the integral as
\begin{linenomath*}\begin{equation}\label{eq: 2D_Term2}
\begin{aligned}
\langle \mu W_j W_{j'} \rangle_{\hat{\Omega}}= & \left \langle \sum_{l=1}^{N_l} \, \sum_{k=-N_l}^{N_l}Y_{l}^{k}(\mu, \varphi) v_{l k j}(t)  \sum_{l'=1}^{N_l} \, \sum_{k'=-N_l}^{N_l}Y_{l'}^{k'}(\mu, \varphi) v_{l' k' j'}(t) \right\rangle_{\hat{\Omega}} \\
= & \left\langle \sum_{l=1}^{N_l} \, \sum_{k=-N_l}^{N_l} \, \sum_{l'=1}^{N_l} \, \sum_{k'=-N_l}^{N_l}
v_{l k j}(t) \, \mu Y_l^k(\mu, \varphi) Y_{l'}^{k'}(\mu, \varphi) \, v_{l' k' j'}(t)  \right\rangle_{\hat{\Omega}} \\
= &  \sum_{l=1}^{N_l} \, \sum_{k=-N_l}^{N_l} \, \sum_{l'=1}^{N_l} \, \sum_{k'=-N_l}^{N_l}
v_{l k j}(t) \left \langle \mu Y_l^k(\mu, \varphi) Y_{l'}^{k'}(\mu, \varphi) \right \rangle_{\hat{\Omega}} v_{l' k' j'}(t)
\end{aligned}
\end{equation}\end{linenomath*}
There are two ways to calculate this term. One is to precompute the time-independent part
$\langle \mu Y_l^k(\mu, \varphi) Y_{l'}^{k'}(\mu, \varphi) \rangle_{\hat{\Omega}}$ which forms a $n \times n$ matrix $C$. Thus Eqs.~\eqref{eq: 2D_Term2} requires $\mathcal{O}(n^2 r)$ operations, which is affordable because usually $n$ is not large and $C$ is sparse. Alternatively, we could calculate $ \langle \mu W_j W_{j'} \rangle_\mu$ on-the-fly by choosing $\mathcal{O}(n)$ quadrature points in angle, and it requires $\mathcal{O}(n r^2)$ operations for all the $r^2$ entries. In this work we chose the first method which is more straightforward to implement.

\subsection{Upwind scheme with slope reconstruction}
We apply the standard upwinding technique to solve the advection equation Eqs.~\eqref{eq: 2D_k_eq}. For the space derivative term we use upwinding to write
\begin{linenomath*}\begin{equation}\label{eq: 2D_dK}
\begin{aligned}
\partial_z K_{j} \, \langle \mu W_j W_{l}\rangle_{\hat{\Omega}} =&\frac{1}{\Delta z}(K_{p, q+\frac{1}{2},j} - K_{p,q-\frac{1}{2},j}) \langle \mu W_j W_{l}\rangle_{\hat{\Omega}}\\
=&\frac{K_{p,q+1,j} - K_{p,q-1,j}}{2\Delta z} \, (V^T C V)_{jl}- \frac{K_{p,q+1,j}- K_{p,q,j} + K_{p,q-1,j}}{2\Delta z} \, (V^T \Sigma V)_{jl} \\
=& \frac{1}{\sqrt{\Delta x \Delta z}} \sum_{i=1}^r  \sum_{p=1}^{N_x} \sum_{q=1}^{N_z} \Bigg ( \left ( \frac{u_{p,q+1,i} - u_{p,q-1,i}}{2 \Delta z} \right ) S_{ij} (V^T C V)_{jl} \\  &- \left ( \frac{u_{p,q+1,i} + u_{p,q-1,i} - 2u_{p,q,i}}{2 \Delta z} \right ) S_{ij} (V^T \Sigma V)_{jl} \Bigg )
\end{aligned}
\end{equation}\end{linenomath*} 
where $\Sigma$ is a stabilization matrix that we take to be a diagonal matrix with the singular values of $C$. Other stabilization terms could be used, including Lax-Friedrichs where $V^T \Sigma V$ is replaced by a constant times an identity matrix. This method allows the spatial variables to be differentiated separately, so the same treatment can be applied to $\partial_x K_{l} \, \langle {\sqrt{1 - \mu^2} \cos \varphi} \, W_j W_{l}\rangle_{\hat{\Omega}}$. 

Additionally, the harmonic mean limiter is adopted to reconstruct this slope term $K_j$ for  better accuracy. We define the slope of $z$ direction in the edge of the cell $(p, q)$ as 
\begin{linenomath*}\[m_{p,q}^+ = \frac{K_{p,q,j}-K_{p,q-1,j}}{\Delta z}, \quad \quad m_{p,q}^- = \frac{K_{p,q+1,j}-K_{p,q,j}}{\Delta z}. \]\end{linenomath*} 
Then the slope of this cell is 
\begin{linenomath*}\begin{equation}
  m_{p,q} =
    \begin{cases}
      \frac{2 m_{p,q}^+ m_{p,q}^-}{m_{p,q}^+ + m_{p,q}^-} & \text{if $m_{p,q}^+ m_{p,q}^- > 0$}\\
      0 & \text{otherwise}
    \end{cases}       
\end{equation}\end{linenomath*}
With the reconstructed edge value then
\begin{linenomath*}\[K_{p,q,j}^+ = K_{p,q,j} + \frac{\Delta z}{2} m_{p,q}, \quad \quad K_{p,q,j}^- = K_{p,q,j} - \frac{\Delta z}{2} m_{p,q}, \]\end{linenomath*} 
Eqs.~\eqref{eq: 2D_dK} can be written as
\begin{linenomath*}\begin{equation}\label{Term2_2nd}
\begin{aligned}
\partial_z K_{j} \, \langle \mu W_j W_{l}\rangle_{\hat{\Omega}} = & \frac{1}{\Delta z}(K_{p, q+\frac{1}{2},j} - K_{p,q-\frac{1}{2},j}) \langle \mu W_j W_{l}\rangle_{\hat{\Omega}} \\
&=\frac{1}{2\Delta z}\Big ((K_{p,q+1,j}^-+K_{p,q,j}^+ - K_{p,q-1,j}^+ - K_{p,q,j}^-)(V^T C V)_{jl}\\
&- (K_{p,q+1,j}^- - K_{p,q,j}^+ - K_{p,q,j}^- + K_{p,q-1,j}^+)(V^T \Sigma V)_{jl} \Big ),
\end{aligned}
\end{equation}\end{linenomath*} 

Spherical harmonics expansions can yield oscillatory or negative solutions \cite{Mcclarren2008}. To address this issue we implemented angular filtering \cite{McClarren2010, Radice2013, Laboure:2016hm} which can significantly increase the performance of $P_n$ method in solving radiative transfer equation by removing the oscillations. We implemented the Lanczos filter into our explicit solver by using an equivalent equation approach \cite{Radice2013} and combined it with the low-rank approximation algorithm. The filtered equation has the form of 
\begin{linenomath*} \begin{equation} \label{Filtered_RadiativeTransfer} 
\begin{aligned}
\frac{1}{c} \frac{\partial \psi (\vec{r}, \hat{\Omega}, t)}{\partial t} + \hat{\Omega} \vd \grad \psi (\vec{r}, \hat{\Omega}, t)  +  \sigmat(\vec{r}) \psi (\vec{r}, \hat{\Omega}, t) + \beta \sigmaf(\vec{r}, \hat{\Omega}) \psi(\vec{r}, \hat{\Omega}, t)
\\
=  \frac{1}{4\pi} \sigmas(\vec{r}) \phi(\vec{r}, t) + Q(\vec{r},t).
\end{aligned}
\end{equation}\end{linenomath*} 
where the free parameter $\beta$ is the filter strength, $\sigmaf = \log \ \frac{\sin \eta}{ \eta} \frac{l}{l+1}$ and $\eta$ is the order of the \PN\ expansion and $l$ is the index of \PN\ moments.

\subsection{Memory Reduction and Computational Efficiency}
The memory footprint required to compute the solution is based on storing the matrices $U$, $V$, and $S$. Therefore, the memory required is $2 (mr+r^2+nr)$ where the factor 2 assumes that we need to store the previous step solution as well as the new step. The full solution to this problem without splitting would require a memory footprint of $2mn$. Therefore, for $r \ll m,n$, there will be large memory savings. 

{Given that double-precision floating-point format occupying 8 bytes in computer memory and $1\,\,\mathrm{megabyte}\, (\mathrm{MB})  = 10^6 \,\, \mathrm{megabytes}$, then the required memory for the low-rank algorithm measured in MB is }
\begin{linenomath*} \begin{equation}\label{eq: 2D_lowrank_Memory}{
\mathrm{memory} = 16 \times (mr+r^2+nr) \times 10^{-6}},
\end{equation}\end{linenomath*} 
{and for the full solution the memory required is}
\begin{linenomath*} \begin{equation}\label{eq: 2D_full_Memory}
{\mathrm{memory} = 16 \times mn \times 10^{-6}}.
\end{equation}\end{linenomath*}

{The low-rank scheme reduces computational cost significantly as well. It is straightforward that solving equation {\eqref{eq: 2D_k_eq}} requires $\mathcal{O}(mr^2)$ operations. Note that the computation of the $\langle \mu W_j W_{j'} \rangle_{\hat{\Omega}}$ term in Eq.~{\eqref{eq: 2D_Term2}} takes $\mathcal{O}(n r^2)$ operations. Equation {\eqref{eq: 2D_s_eq}} also needs $\mathcal{O}(mr^2)$ operations where the term {\eqref{eq: 2D_Term1}} is the heaviest computational burden. Since $X$ is preserved in equation {\eqref{eq: 2D_s_eq}} and {\eqref{eq: 2D_L_eq}}, there is no need to compute term {\eqref{eq: 2D_Term1}} again in the third equation. Therefore, equation {\eqref{eq: 2D_L_eq}} only requires $\mathcal{O}(nr^2)$ operations. The two QR decompositions consume another $\mathcal{O}(mr^2) + \mathcal{O}(nr^2)$. Combining all of these costs, in total the low-rank scheme requires $\mathcal{O}(mr^2) + \mathcal{O}(nr^2)$ operations whereas the full solution needs $\mathcal{O}(mn^2)$.}

{We note that we have not considered the applications of the above algorithm on distributed-memory parallel computation. In such a scenario the projection operators, and the procedures required to obtain them, may hurt scalability and performance when applied globally. We contend that applying the low-rank technique on individual processing elements (PEs) would alleviate this shortcoming.  The full solution would only be required on domain boundaries, and each PE would be able to have its own low-rank structure, possibly improving the results.  A study of such a implementation is outside the scope of this paper and will be the focus of future research. }

\subsection{Conservation}
The low-rank algorithm we have described does not conserve the number of particles. This loss of conservation is a result of information lost in the algorithm when restricting the solution to low-rank descriptions. We have addressed this by globally scaling the solution after each time step to correct for any particles lost.  We can do this because we know the number of particles that are absorbed, travel across the boundary, and born from sources. This correction will not preserve higher moments of the solution, and it only preserves the zeroth-order moment in space and angle. This issue has also been addressed in \cite{Einkemmer2019}, where a correction term calculated by the imposed conservation law is added to each splitting step {to ensure global and/or local conservation of mass and momentum (the zeroth and first moments in our problem) by the solution of a minimization problem. The procedure outlined in {\cite{Einkemmer2019}} could be applied to the radiative transfer equations. We choose the simpler scaling because we are primarily interested global conservation of the zeroth moment as it is through this quantity where the primary coupling to other physics occurs.} 

\section{Numerical Results}
We run the transport simulations with five test problems, including the plane source problem and Reed's problem in 1D slab geometry in addition to the line source problem, lattice problem and {double chevron problem} in 2D planar geometry. {We note that it is common for radiation transport problems to have many more spatial degrees of freedom $m$ than angular degrees of freedom $n$. To test our method under these common circumstances we } fix the spatial resolution and vary $n$ and the maximum rank $r$ in the following simulations. The CFL condition with the formula $CFL = \frac{\Delta t}{\Delta z}$ in 1D and $CFL = \min(\frac{\Delta t}{\Delta x},\frac{\Delta t}{\Delta z})$ in 2D is set to $0.2$, {except for the line source problem where $CFL= 0.1$ }. The conservation fix is applied to the plane and line source problems unless specified otherwise. Additionally, we used one simulation to demonstrate that the low-rank algorithm is compatible with the filter, other than that all the results are unfiltered. Our implementation is written in MATLAB, and our particle speeds are set to $c=1 \,$ cm/s. 

{We use Eqs.~{\eqref{eq: 2D_lowrank_Memory}} or {\eqref{eq: 2D_full_Memory}} to calculate the theoretical value of the memory footprint for the computation. Additionally, we adopt a MATLAB built-in function ``memory" to measure the actual memory usage for the 2D problems. It is evaluated by recording the maximum running memory during the simulation and subtracting the memory occupied by MATLAB before the solution function was called. Along with the memory, we track the execution time with the ``profile'' function in MATLAB. We point out that the recorded memory and running time could vary a small percentage over different simulations. To account for this variation, we take the average values of the memory usage and the running time in 20 simulations for each test. The unit of memory in this section is MB.}

\begin{figure}[h!] 
	\begin{subfigure}{1\textwidth}\caption{Low-rank solutions without scaling}
	\centering
	\includegraphics[scale=0.2]{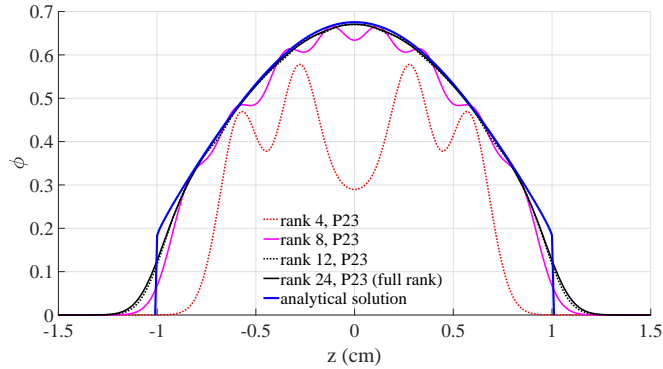}
	\end{subfigure}
	\begin{subfigure}{1\textwidth}\caption{Low-rank solutions with scaling}
	\centering
	\includegraphics[scale=0.2]{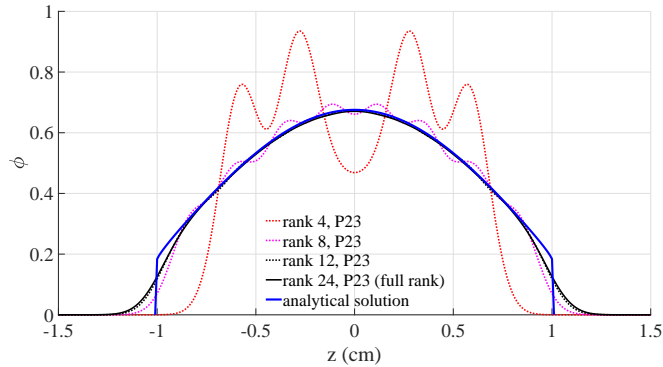}
	\end{subfigure}
	\caption{Solutions to the plane source problem at $t = 1 \,$s using the low rank method compared to the analytic solution. Numerical smearing errors are observed in both edges due to numerical dissipation \cite{Raithby1999}.}
	\label{PP_T1}
\end{figure}

\begin{figure}[h!]
\centering
\includegraphics[scale=0.2]{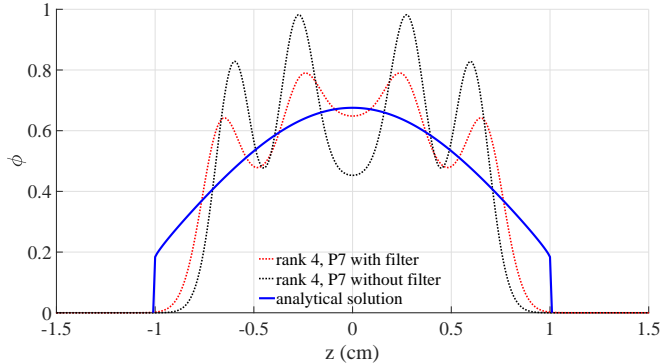} 
\caption{Comparison of P$_7$ solutions of rank $4$ with and without a filter to the analytic solution at $t=1 \,$s.}
\label{PP_filter_T1}
\end{figure}

\begin{figure} 
	\centering
	\includegraphics[scale=0.2]{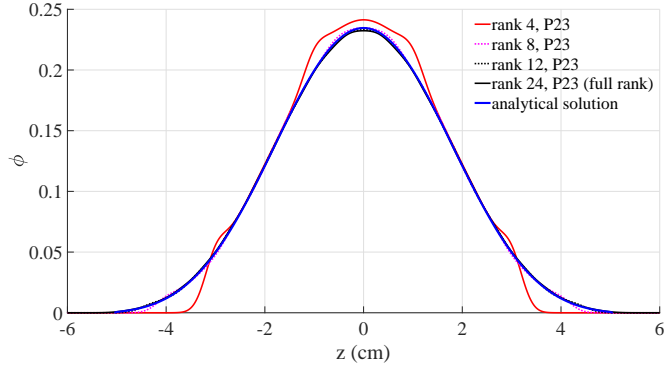} 
	\caption{Solutions to the plane source problem using the low rank method compared to the analytic solution at $t=5 \,$s.}   
	\label{PP_T5}
\end{figure}	

\begin{figure}[h!] 
	\begin{subfigure}{1\textwidth}\caption{The error of low rank solutions at t=1}
	\centering
	\includegraphics[scale=0.2]{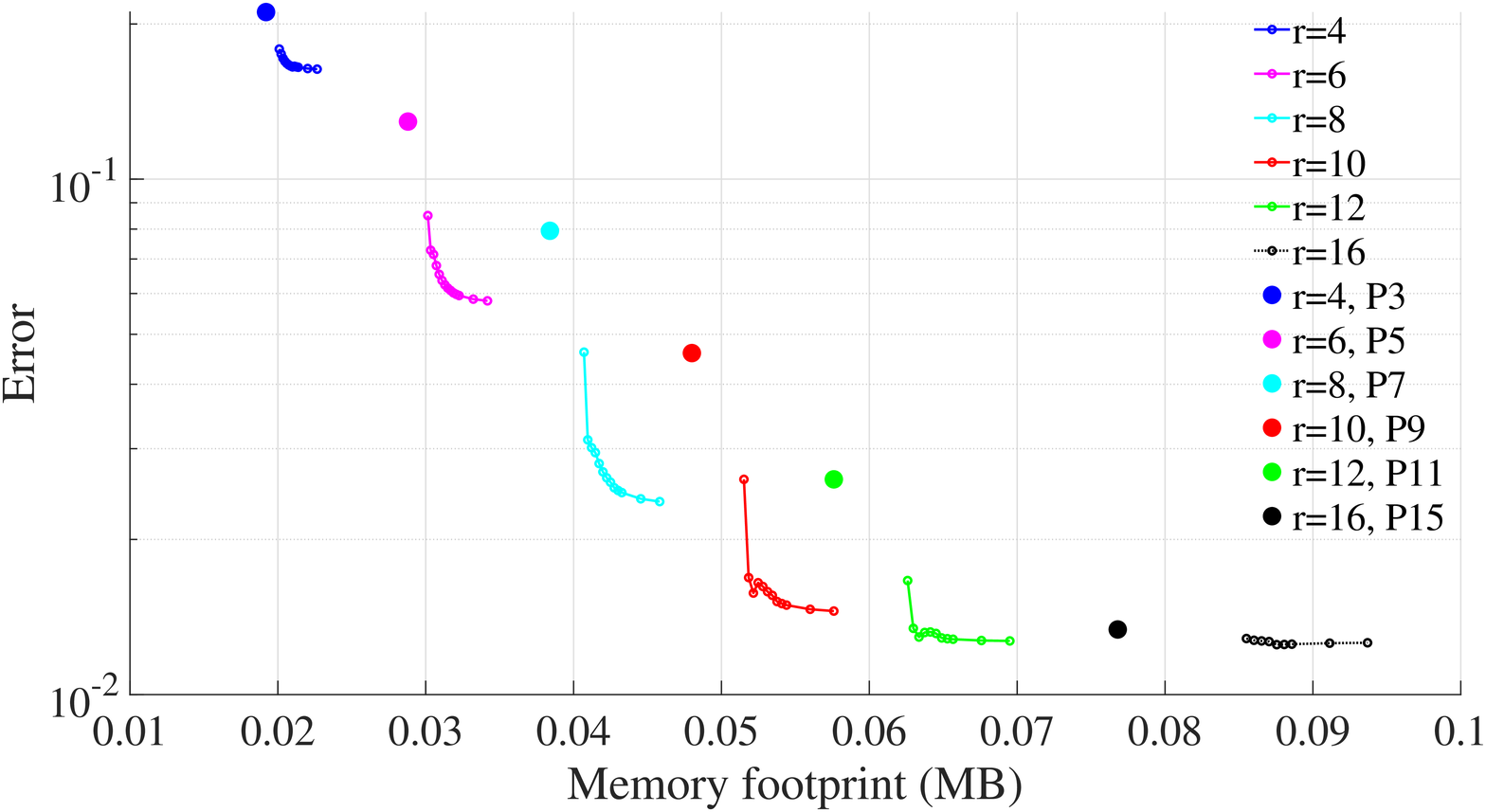}
	\end{subfigure}
	\begin{subfigure}{1\textwidth}\caption{The error of low rank solutions at t=5}
	\centering
	\includegraphics[scale=0.2]{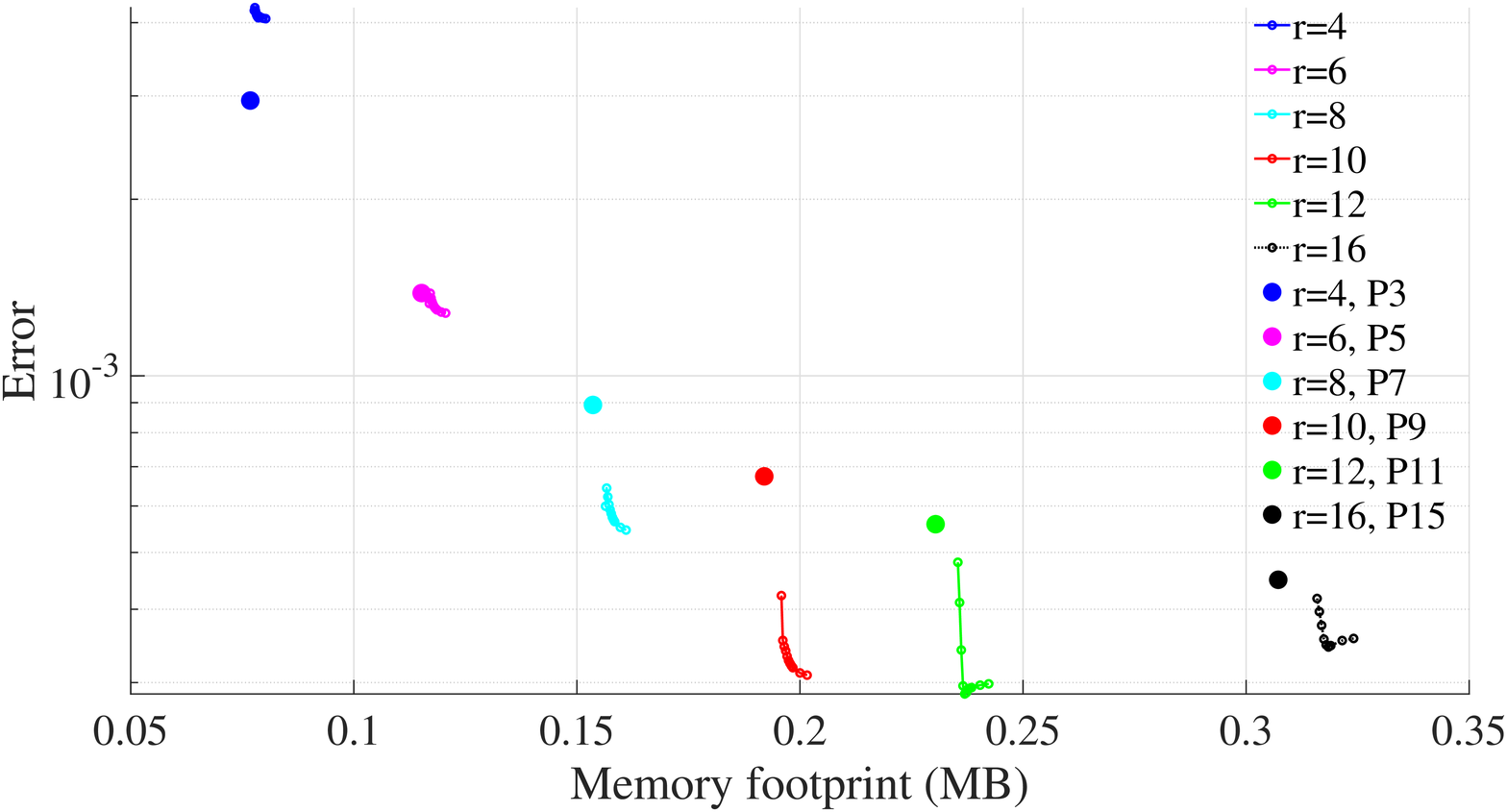}
	\end{subfigure}
	\caption{The comparison of errors on the plane source problem with $m=300$ and with different theoretical memory usage by varying $r$ and $n$ are shown. Each line represents the error with a fixed rank that varies the number of angular basis functions $n$. The bold dot denotes the full rank solution with $n=r$.}
	\label{PP_results}
\end{figure}

\subsection{Plane source problem}
The plane source problem describes a plane of isotropically moving particles  emitted at $t=0$ in a purely scattering medium with no source, i.e., $\sigmat = \sigmas = 1 \,$cm$^{-1}$ and $Q = 0$. The only spatial variable in this slab geometry is $z$; the initial condition is given by a delta function as $\psi(z,\mu,t) = \delta(z)/2$. For the simulation parameters, we set the spatial resolution to be $\Delta z = 0.01$ cm, which corresponds to the number of mesh zones $m = 300$ for the $t=1 \,$s solution and $m = 1200$ for $t=5 \,$s. $P_{23}$ solutions with different rank are compared. When used, the filter strength $\beta$ is set to $50$. The analytical benchmark solution that we compare to was given by Ganapol \cite{Ganapol2008}.

Figure \ref{PP_T1} shows the solutions of varying rank and Legendre polynomial orders with and without the conservation fix where the zeroth moment is scaled. As can be seen in either case, the solution with rank $12$, which is half of the full rank, matches the analytic solution to the scale of the graph in the middle part of the problem. It also agrees well with the full rank solution. The rank $8$ solutions still capture the analytical solution well. However, the loss of conservation can be observed when the conservation fix is not applied. Rank $4$ is not sufficient for accurate results and shows more affect from the loss of conservation. As shown in Figure \ref{PP_filter_T1}, the low-rank solution can be improved by the filter: P$_7$ solutions of reduced rank improve when a filter is used. Figure \ref{PP_T5} presents the solution at $t = 5 \,$s, at which even rank $8$ appears to be sufficient. At this later time, there are few remaining uncollided particles from the initial condition.  Therefore, fewer angular degrees of freedom is needed.
    
For a more quantitative comparison, the root mean square (RMS) error of the numerical results with different $n$ and $r$ is shown in Figure \ref{PP_results}. In this figure, the colors for the symbols and lines correspond to the rank used in a calculation, and different values of the $n$, the number of angular basis functions, are corresponding dots. For each color the value of $n$ ranges from $r$ to $100$. The large points are the value of the error using the standard full rank method with $r=n$. We can observe that the low-rank solution is more accurate than the full rank with the same memory usage. For example, the error of full rank solution $n=12$ at $t = 1 \,$s using with a memory footprint of $0.057$ MB is about $0.026$. With the same memory, the error can be reduced to $0.013$. We can also use $70\%$ of the memory to achieve the same accuracy as seen in the $r=8$ error. Increasing the resolution and rank will contribute to the accuracy of solutions. Given the way we performed this study with a fixed spatial mesh and time step and the conservation fix we used, we can see some error stagnation in the low-rank solution at $t=5 \,$s. Other numerical experiments indicate that increasing the number of spatial zones can further decrease the error.

\begin{figure} 
	\centering
	\includegraphics[scale=0.4]{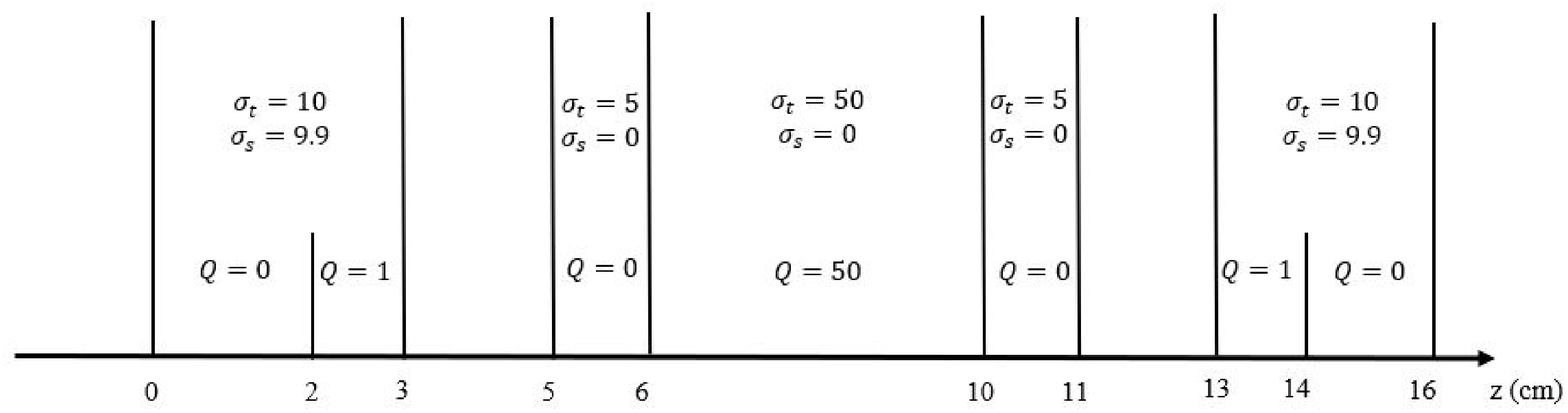}
	\caption{The material layout in Reed's problem where the blank zone means vacuum and the unit of cross sections is cm$^{-1}$.}
	\label{Reeds_layout}
\end{figure}

\begin{figure}[h!] 
	\begin{subfigure}{1\textwidth}
	\centering
	\includegraphics[scale=0.2]{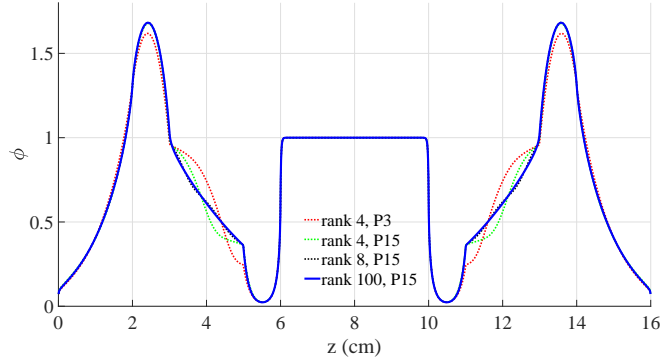}
	\end{subfigure}
	\caption{Solutions to the Reed's problem at $t=5 \,s$ using the low rank method compared to the high-degree full rank solution. The rank 8 $P_{15}$ solutions are mostly obscured by the $P_{99}$ solution in this plot.}
	\label{Reeds_results}
\end{figure}

\begin{figure}[h!] 
	\begin{subfigure}{1\textwidth}
	\centering
	\includegraphics[scale=0.2]{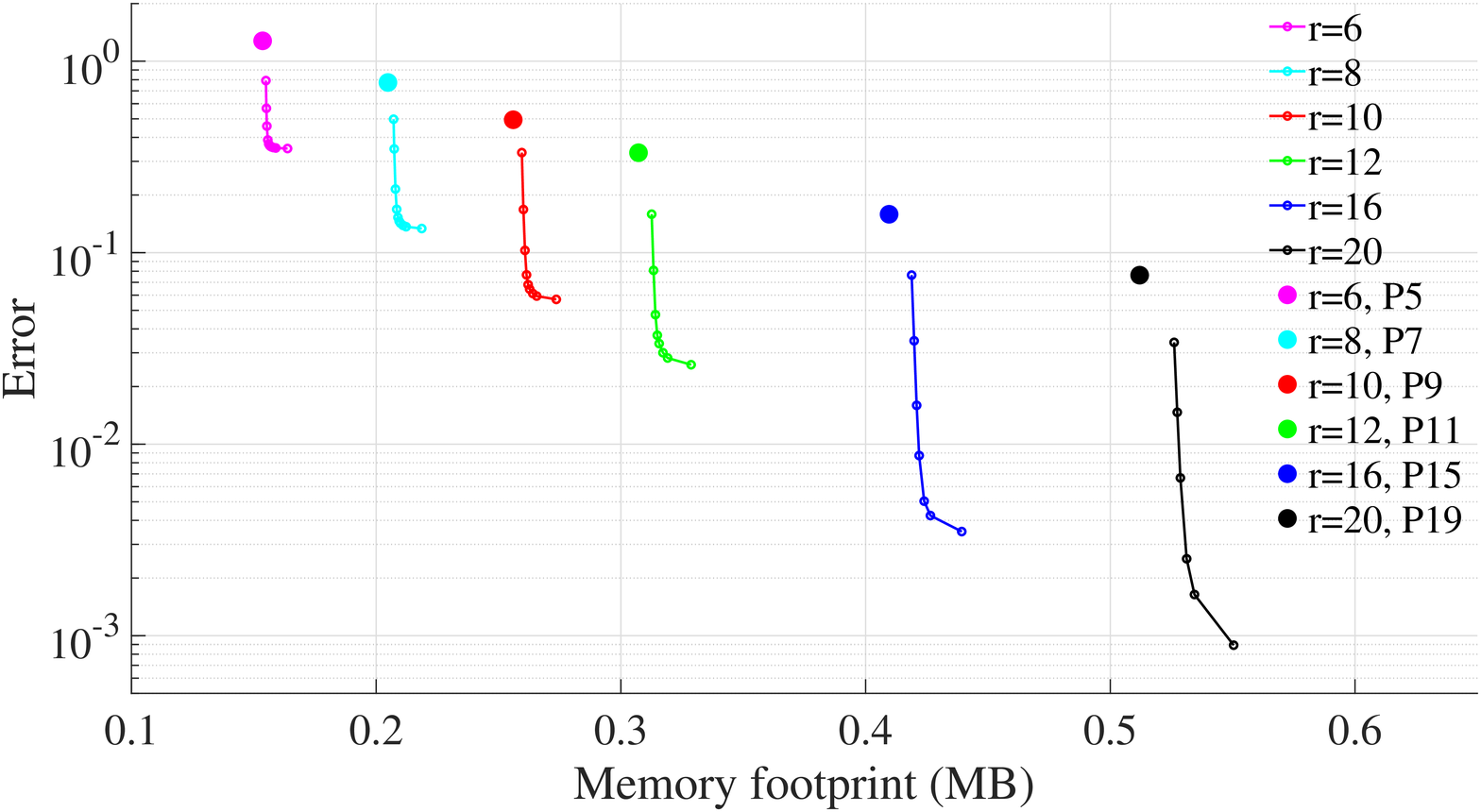}
	\end{subfigure}
	\caption{The comparison of errors for Reed's problem with $m=1600$ and with different values for $r$ and $n$. Each line represents the error with a fixed rank $r$ that varies the number of angular basis functions $n$. The bold dot denotes the full rank solution.}
	\label{Reeds_error}
\end{figure}

\subsection{Reed's problem}
The second test problem is Reed's problem, which is a multi-material problem, and its set-up is detailed in Figure \ref{Reeds_layout}. Because Reed's problem does not have an analytical solution, a numerical result with a high degree of angular basis and full rank, where $\Delta z = 0.01$ cm and $P_{99}$ (corresponding to $m=1600$ and $n=100$) is set as a benchmark for memory analysis. Figure \ref{Reeds_results} shows that the rank $4$ solutions differ in the vacuum regions ($z \in (3,5)$ and $z \in (11,13)$) and the scattering region with source ($z \in (2,3)$ and $z \in (13,14)$), but the rank $8$ solution matches the P$_{99}$ reference solution well.  

It can be observed in Figure \ref{Reeds_error} that the low-rank solutions (solid lines with small dots) can give solutions with comparable errors to the full rank solutions (large dots) with much larger memory. For example, the rank $10$ solutions obtain a solution error better than the full rank P$_{19}$ solution with less memory.

\begin{figure}[h!] 
	\begin{subfigure}{.5\textwidth}\centering
	\caption{Analytical solution}
	\includegraphics[width=\columnwidth]{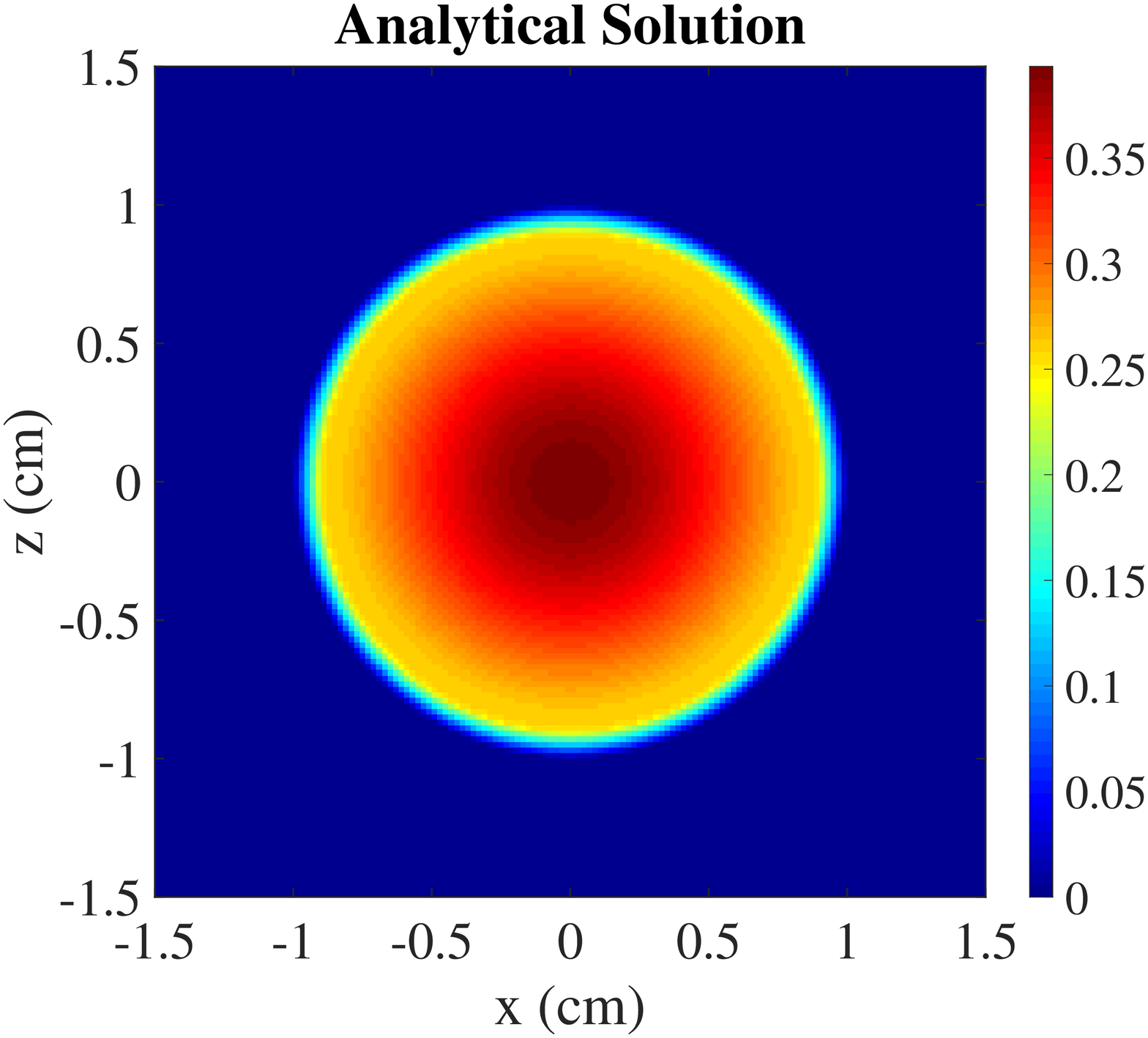}
	\end{subfigure}
	\hfill
	\begin{subfigure}{.5\textwidth}\centering
	\caption{Rank 210, P19}
	\includegraphics[width=\columnwidth]{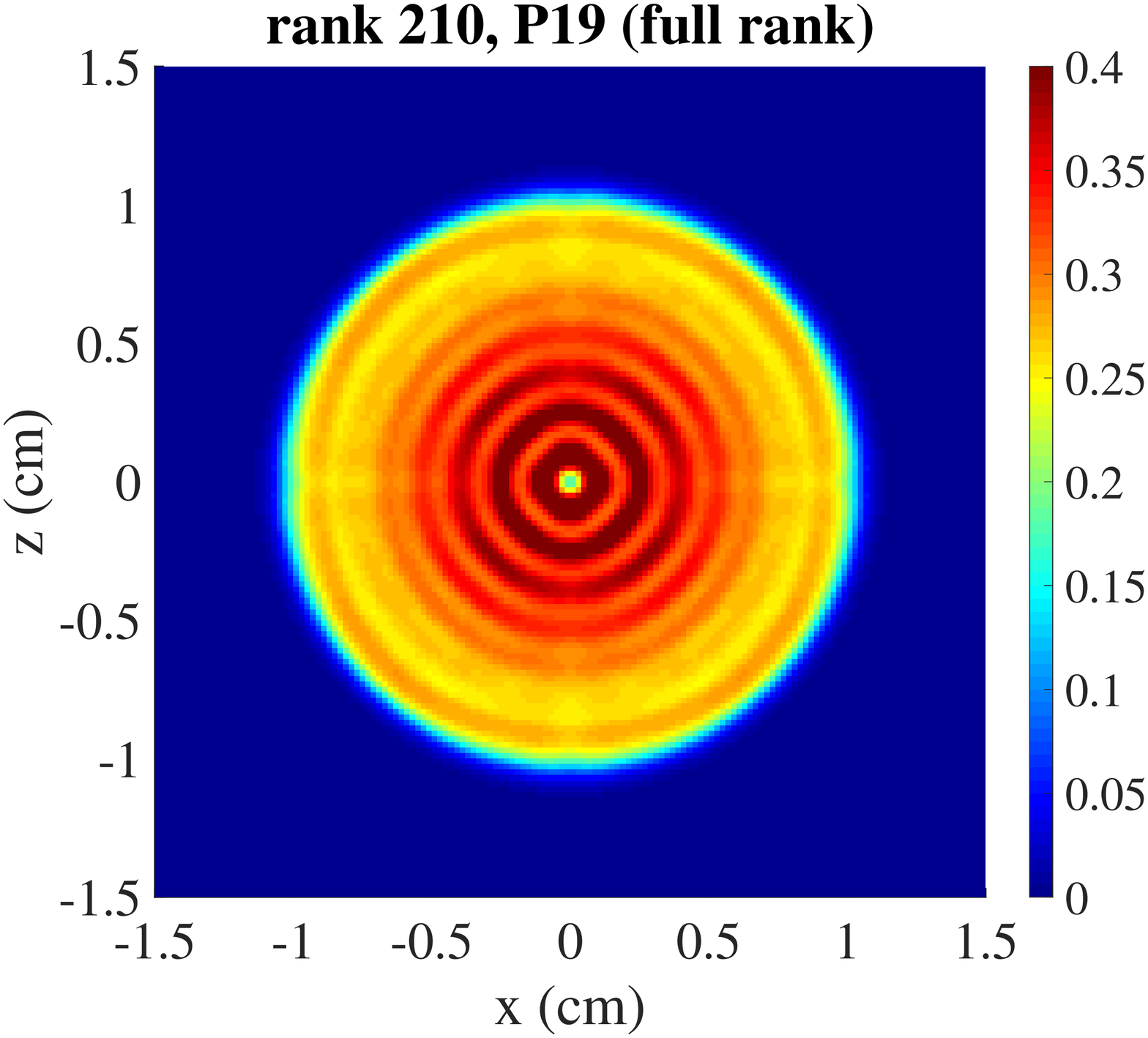}
	\end{subfigure}
	\begin{subfigure}{.5\textwidth}\centering
	\caption{Rank 210, P29}
	\includegraphics[width=\columnwidth]{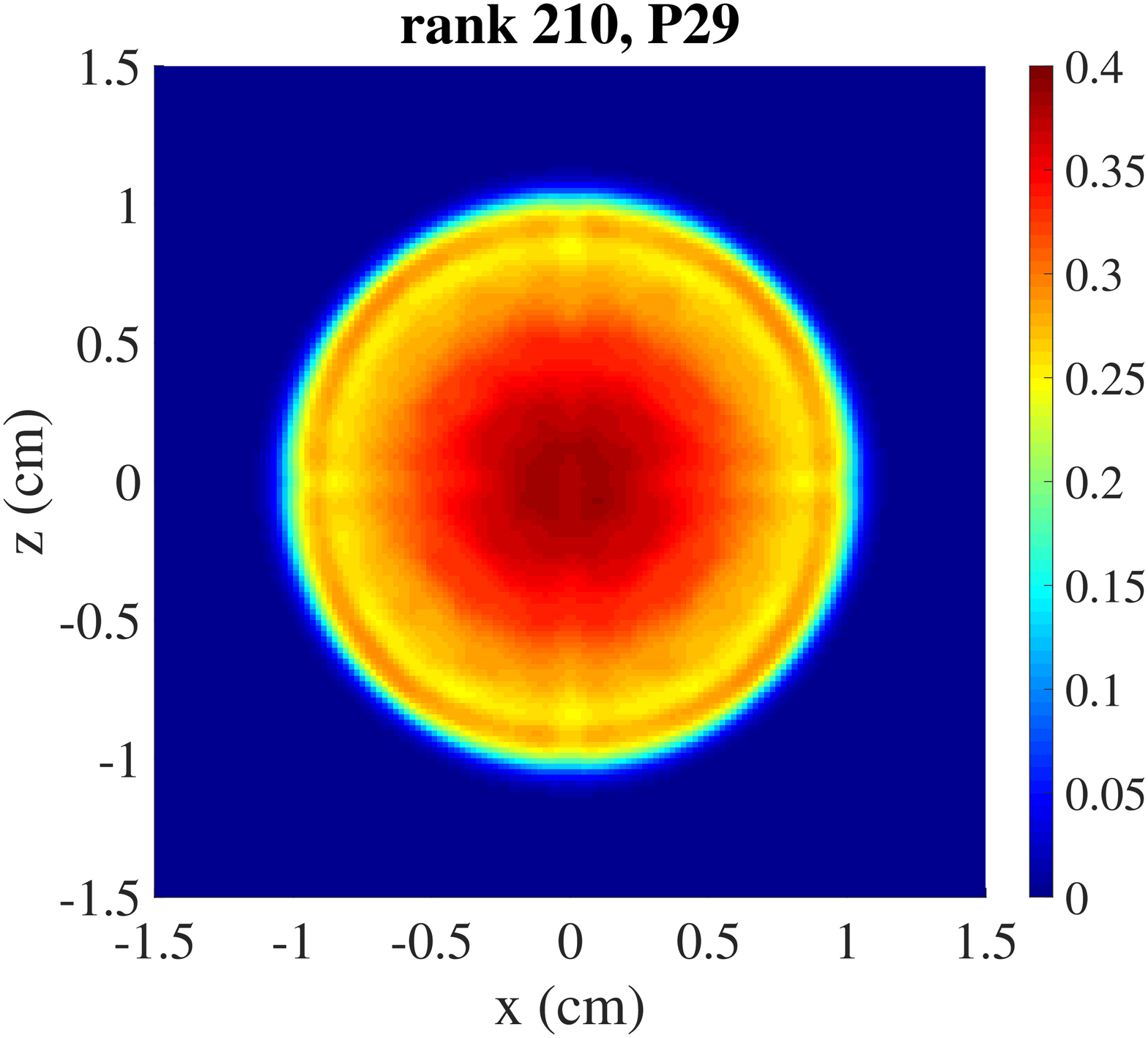}
	\end{subfigure}
	\hfill
	\begin{subfigure}{.5\textwidth}\centering
	\caption{Rank 210, P39}
	\includegraphics[width=\columnwidth]{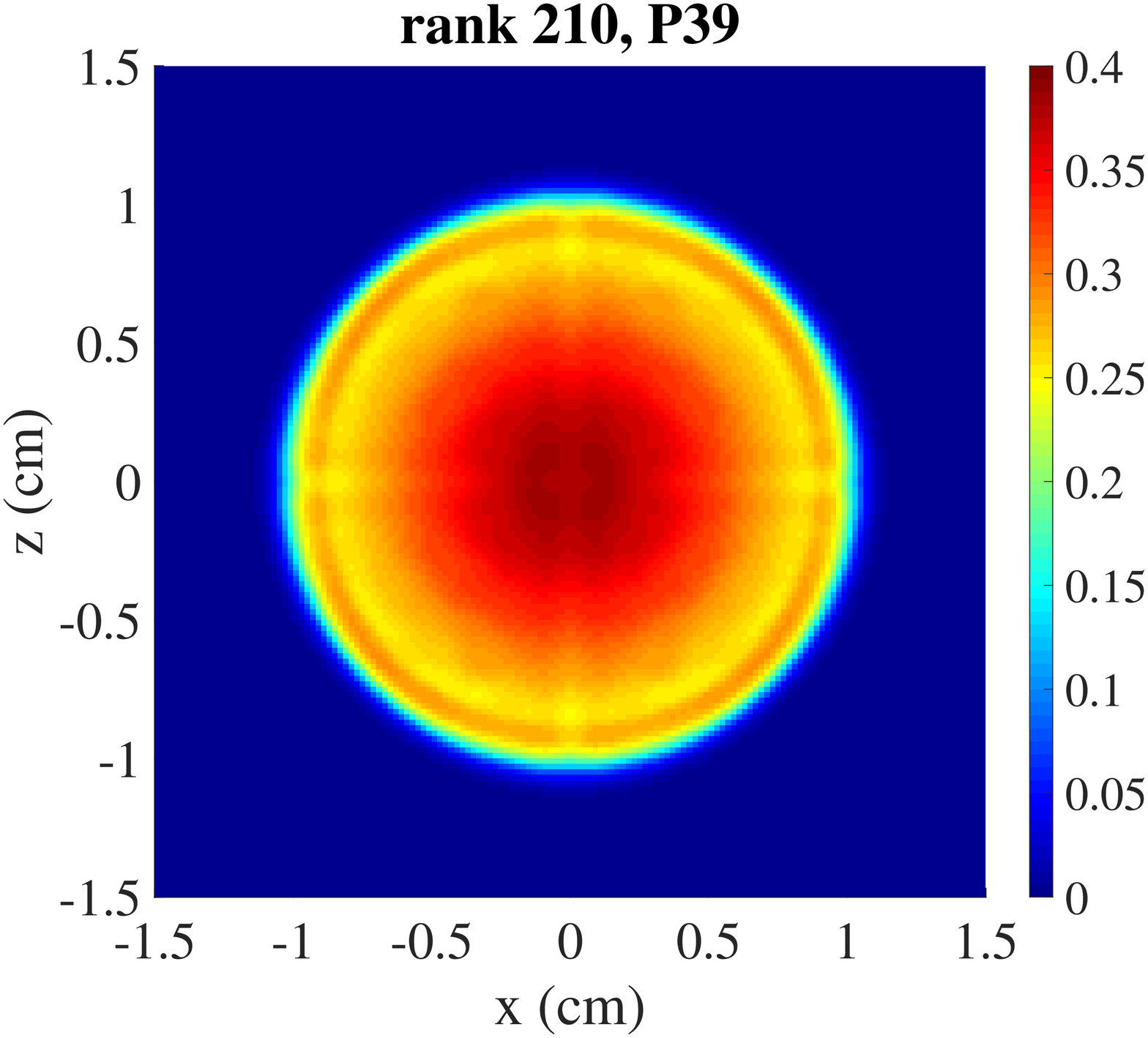}
	\end{subfigure}
	\begin{subfigure}{1\textwidth}\caption{Solution at $z=0$}
	\centering
	\includegraphics[scale=0.2]{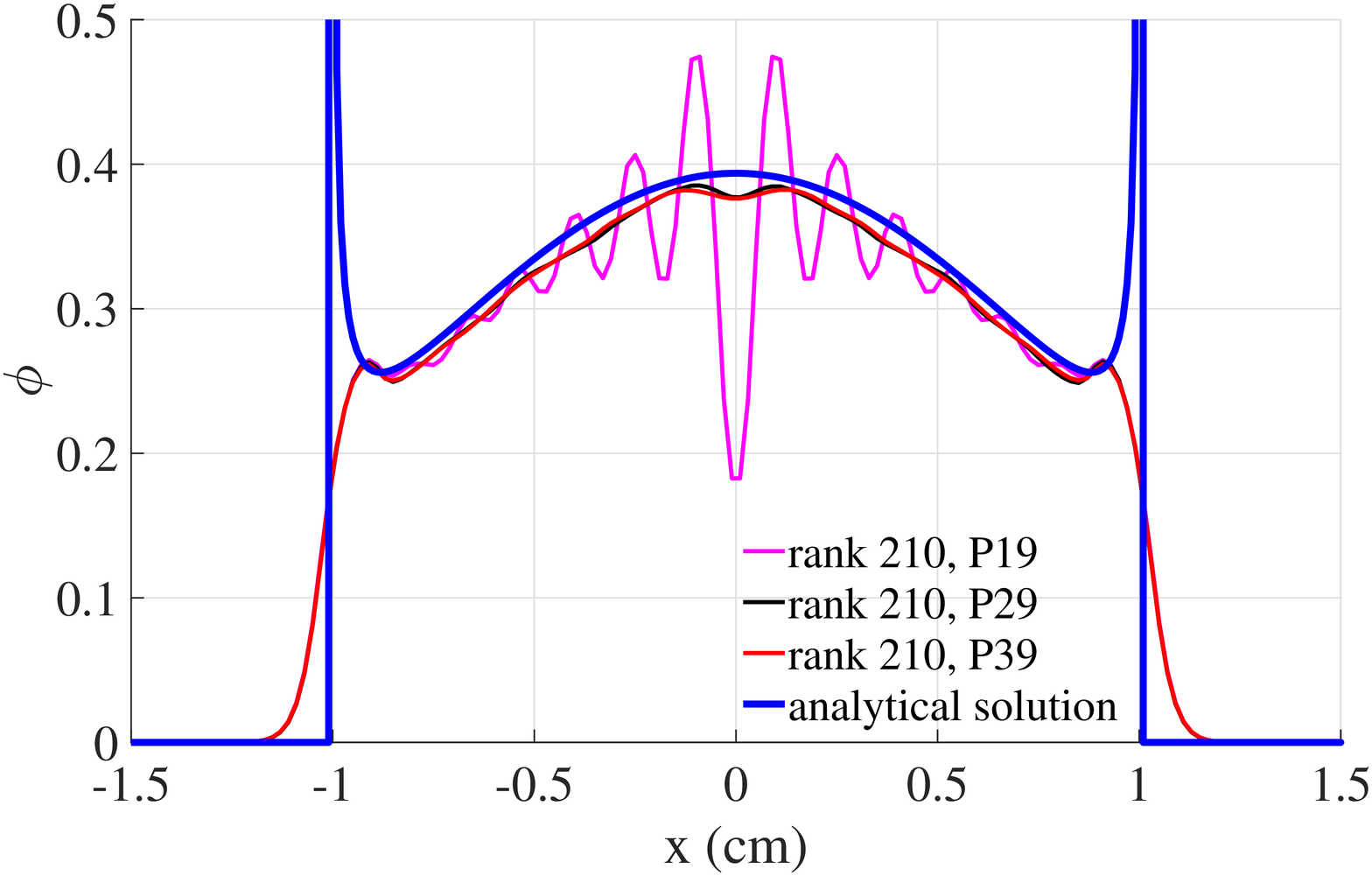}
	\end{subfigure}
	\caption{Solutions to the line source problem at $t = 1 \,s$ using rank $210$ compared to the analytic solution. }
	\label{LS_results1}
\end{figure}

\begin{figure}[h!]
	\begin{subfigure}{1\textwidth}
	\centering
	\includegraphics[scale=0.2]{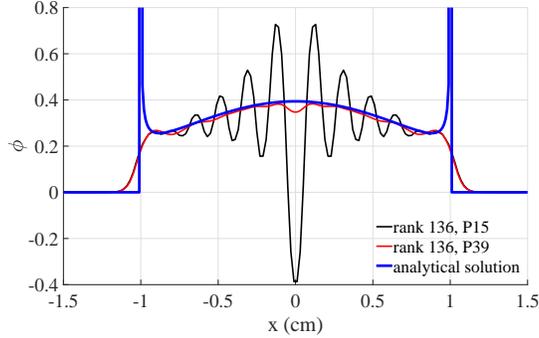}
	\end{subfigure}
	\caption{Solutions to the line source problem at $t = 1 \,s$ using rank $136$ compared to the analytic solution at the cut $z=0$.}
	\label{LS_results2}
\end{figure}

\begin{figure}[h!]
	\centering
	\includegraphics[scale=0.2]{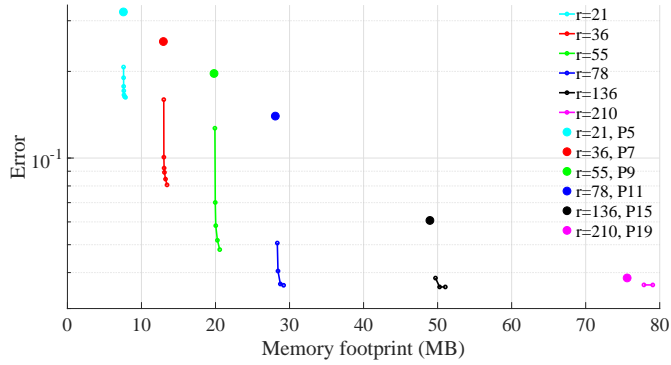} 
	\caption{The comparison of errors for the line source problem with different memory usage are shown. Each dotted line represents the error with a fixed rank that varies the number of angular basis functions $n$. The bold dot denotes the full rank solution.}
	\label{LS_error}
\end{figure}

\begin{figure}[h!]
	\centering
	\includegraphics[scale=0.2]{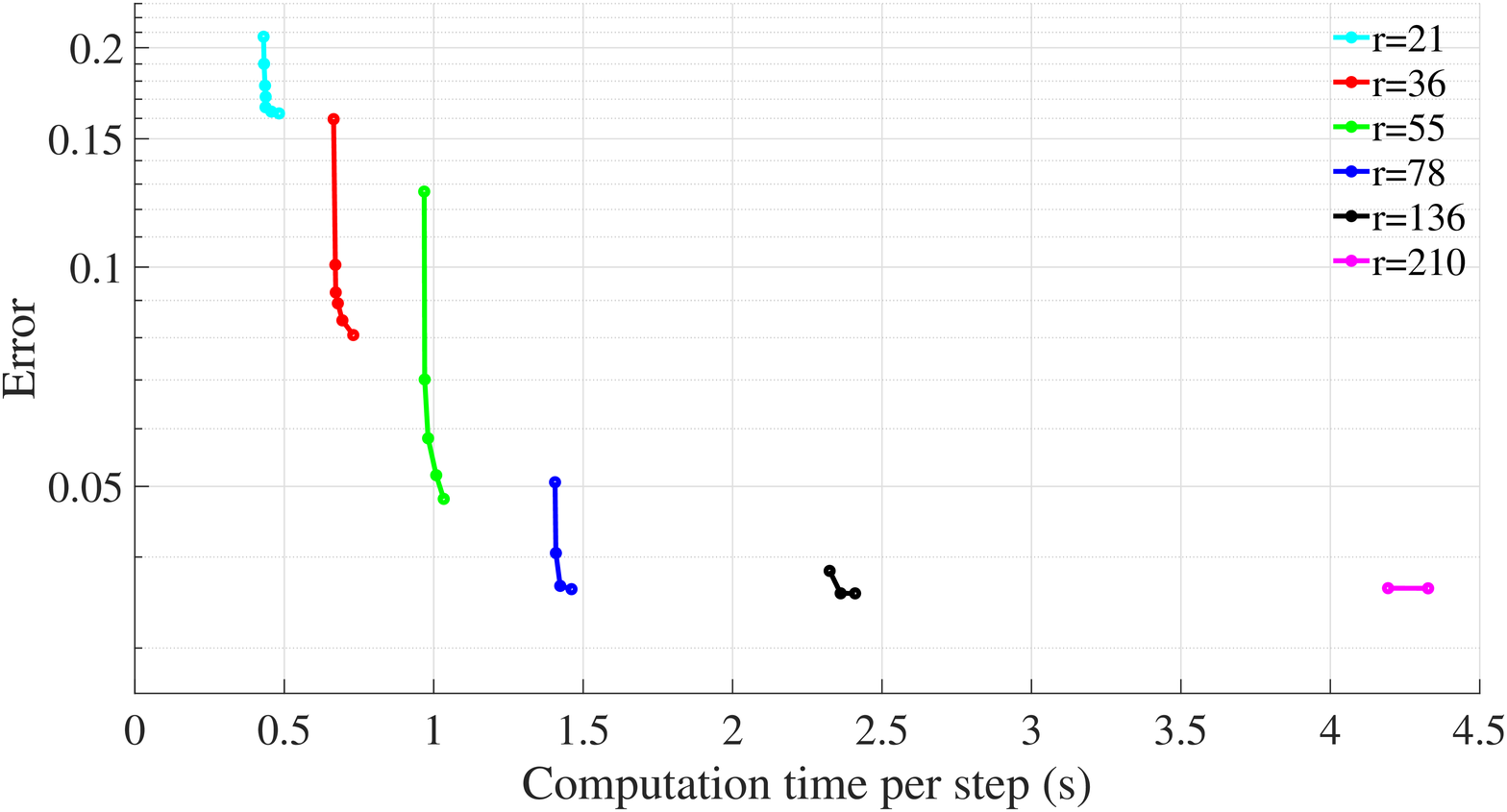} 
	\caption{{The comparison of errors for the low rank solutions of line source problem with different computation time are shown. Each dotted line represents the error with a fixed rank that varies the number of angular basis functions $n$.}}
	\label{LS_error_time}
\end{figure}

\subsection{Line source problem}
The line source problem is a natural extension of the plane source problem to 2-D, where the plane source becomes a line. In this problem we have $\sigmat = \sigmas = 1 \, $cm$^{-1}$, $Q = 0$ and the initial condition $\psi(x, z, \mu, \varphi, t) = \delta(x)\delta(z)/4\pi$. We use a computational domain of $[-1.5,1.5] \times [-1.5,1.5]$ for the simulation time $t=1 \,$s, while the spatial grid is set to be $150 \times 150$. 

The analytical solution from Ganapol \cite{Ganapol2008} and the rank 210 solutions with P$_{19}$, P$_{29}$ and P$_{39}$ are shown in Figure \ref{LS_results1}, from which we can see that the P$_{15}$ solution which is full rank, has large oscillations. The results obtained with the same rank but more angular basis functions are much better, as we can see in Figure \ref{LS_results1}e where fixing the rank and increasing the expansion order improves the solution dramatically. Figure \ref{LS_results2} demonstrates that increasing the angular resolution while fixing the rank improves the solution considerably. 
In this figure, we note that the full rank P$_{15}$ solution has negative values, which are due to the oscillatory property of the \PN\ method \cite{Mcclarren2008}. This nonphysical phenomenon can be alleviated by increasing the order of \PN; using the low-rank approximation this requires only a modest increase in memory. 

Figure \ref{LS_error} shows that memory saving is more significant in the 2D problem, where the number of spatial and angular degrees of freedom is large. As we can see, the theoretical memory footprint is a strong function of rank. It indicates that the accuracy can be improved as we increase the P$_N$ order with modest additional cost. For example, the error of the solution at rank 78 with the low-rank method is four times smaller than the full rank solution. 

{For this problem we measured the memory usage while the program was running to compare it to the theoretical values. We also measured the time required per time step. These results, as well as the theoretical memory usage and RMS error, are shown in Figure {\ref{LS_error_time}} and Table {\ref{linesource_comparsion}}. The implementation used for the full-rank results is a standalone code that does not perform any of the projections or other special steps required in the low-rank algorithm. From this table we see that our MATLAB implementation is using memory very close to the theoretical values. Due to using sparse matrices where possible in both implementations, the actual memory used can be less than the theoretical values. Furthermore, as we shall discuss more later, the actual memory used is a linear function of the theoretical size, indicating that the theoretical value is a reasonable measure of the memory required.} 

{As can be seen from Figure $\ref{LS_error_time}$, the running time of the low-rank solutions depends strongly on rank as well. We point out that the shape of each line in this figure is similar to the line with same rank in Figure $\ref{LS_error}$. This implies that increasing the P$_N$ order in low-rank solutions with the rank fixed will only result in a slight growth in computational cost, when $m \gg n, r$. This is nearly ideal: we decrease memory and computation time by using our low-rank method.}

{From Table {\ref{linesource_comparsion}} we see that the memory and time difference between a full-rank calculation and a low-rank calculation of the same rank and larger $n$ requires is relatively small. However, there is an increase in the running time when switching from the simpler full-rank method to the low-rank method. Nevertheless, when we look at the error per time we see that the low-rank results are significantly better in terms of memory and running time: The low-rank solution of $P_{39}$ with $r=78$ requires an order of magnitude less memory (28.06 MB to 298.88 MB) and is over six times faster (1.46 to 9.79 s) to achieve a lower error (0.0361 to 0.0387).}

\begin{table*}
\caption{\label{linesource_comparsion} The quantitative comparison for the line source problem of memory footprints, {running time} and errors with the same rank but different angular approximations.}
\centering
\begin{tabular}{|c|c|c|c|c|c|}
\hline 
& \makecell{rank 55, $P_{9}$ \\ (full rank) } & \makecell{rank 55, $P_{39}$ \\ (low rank) } & \makecell{rank 78, $P_{11}$ \\ (full rank) } & \makecell{rank 78, $P_{39}$ \\ (low rank) } & \makecell{rank 820, $P_{39}$ \\ (full rank) }\\
\hline
\makecell{Theoretical \\ memory (MB)} & 19.80 & 20.57 & 28.08 & 29.20 & 295.20\\
\hline
\makecell{MATLAB \\ memory (MB)} & 19.69 & 20.24 & 27.98 & 28.06 & 298.88\\
\hline
\makecell{Running time \\ per step (s)} & 0.68 & 1.08 & 0.92 & 1.46 & 9.79\\
\hline
Error & 0.2071 & 0.0481 & 0.1473 & 0.0361 & 0.0387\\
\hline
\end{tabular}
\end{table*}


\begin{figure} 
	\centering
	\includegraphics[scale=0.5]{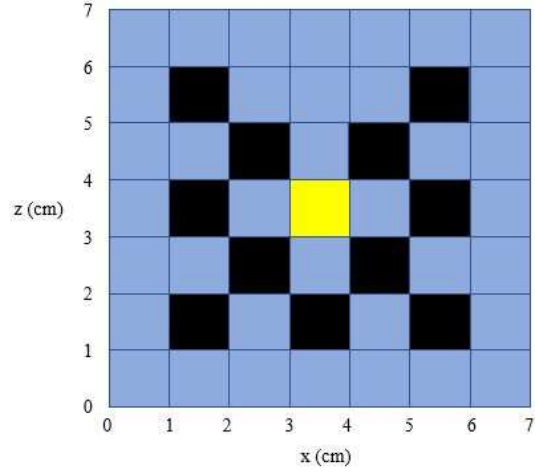}
	\caption{The material layout in the Lattice problem. The blue zones are purely scattering region, the black are absorbing region and the yellow is also the scattering region with an isotropic source which is turned on at t$ = 0$. The checkerboard is surrounded by vacuum.}
	\label{Lattice_layout}
\end{figure}

\begin{figure}[h!] 
	\begin{subfigure}{.5\textwidth}\centering
	\includegraphics[width=\columnwidth]{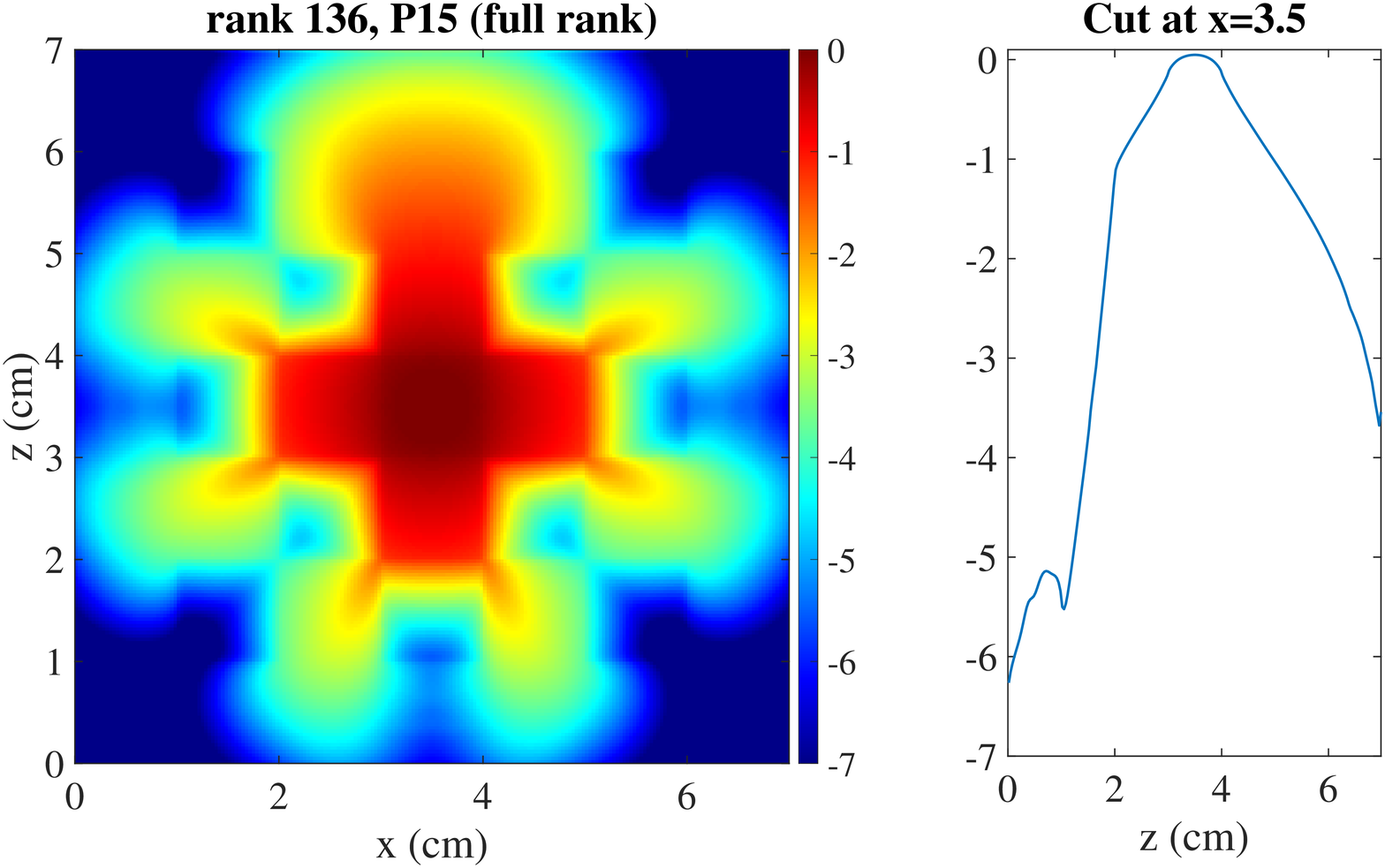}
	\end{subfigure}
	\hfill
	\begin{subfigure}{.5\textwidth}\centering
	\includegraphics[width=\columnwidth]{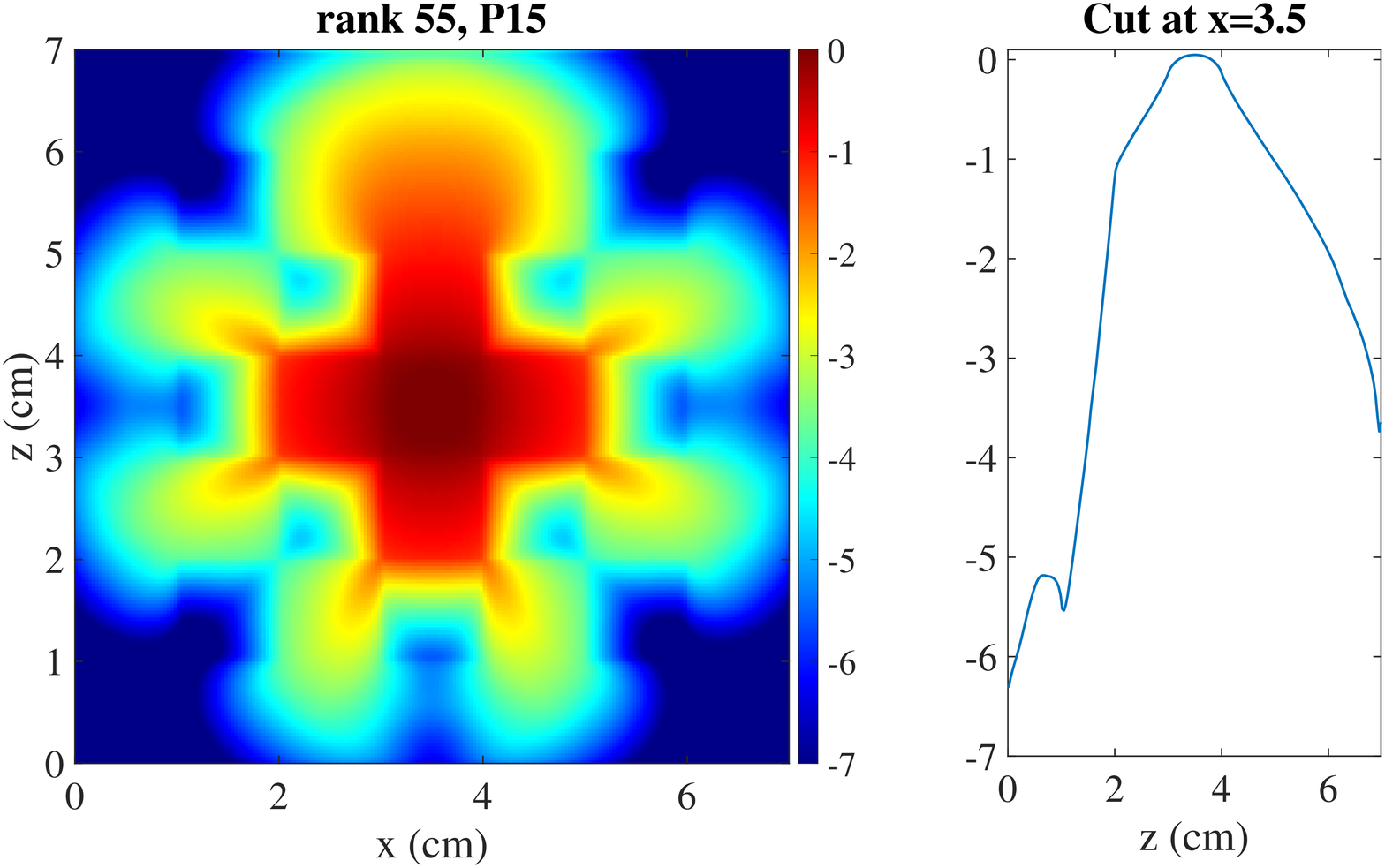}
	\end{subfigure}
	\begin{subfigure}{.5\textwidth}\centering
	\includegraphics[width=\columnwidth]{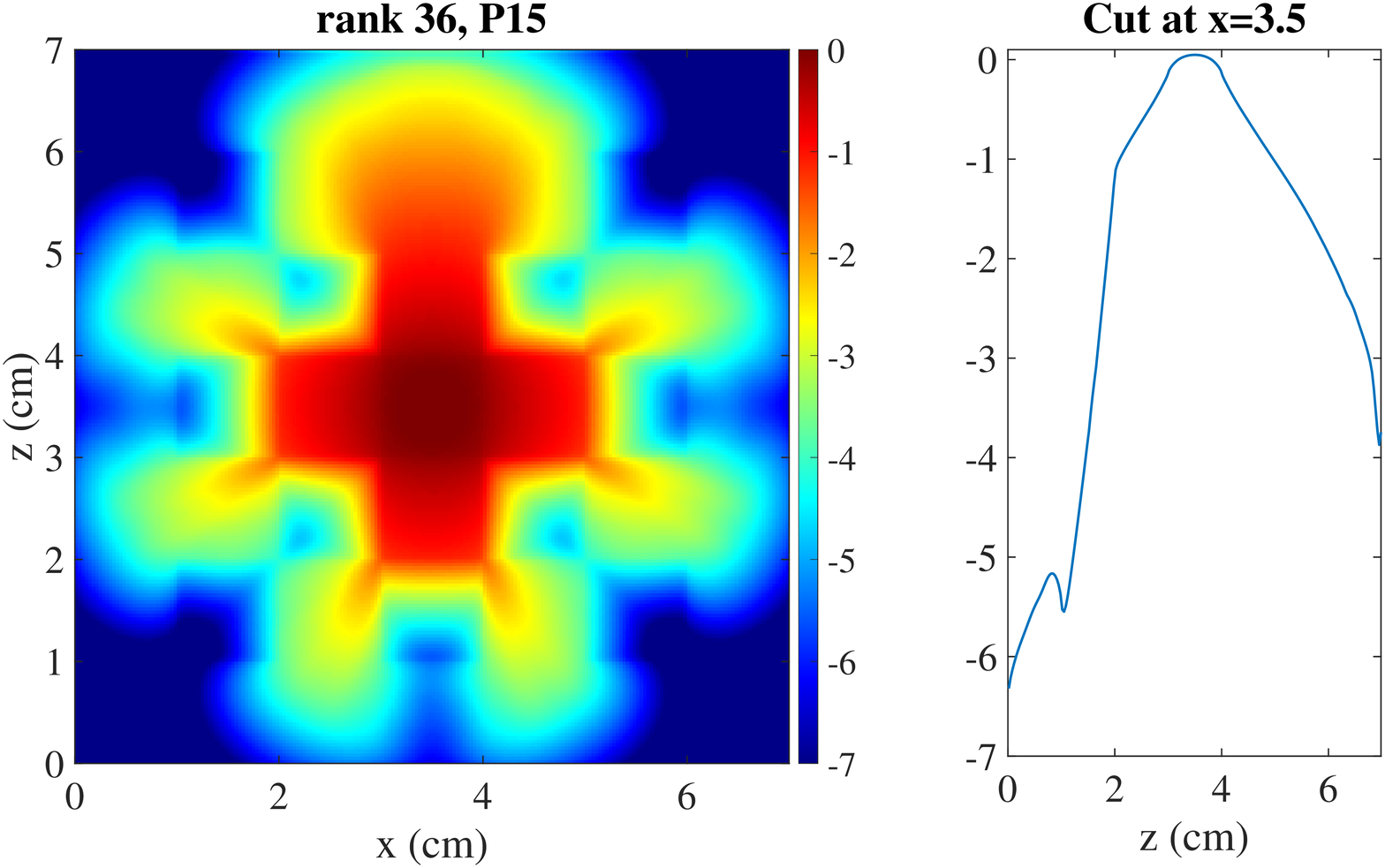}
	\end{subfigure}
	\hfill
	\begin{subfigure}{.5\textwidth}\centering
	\includegraphics[width=\columnwidth]{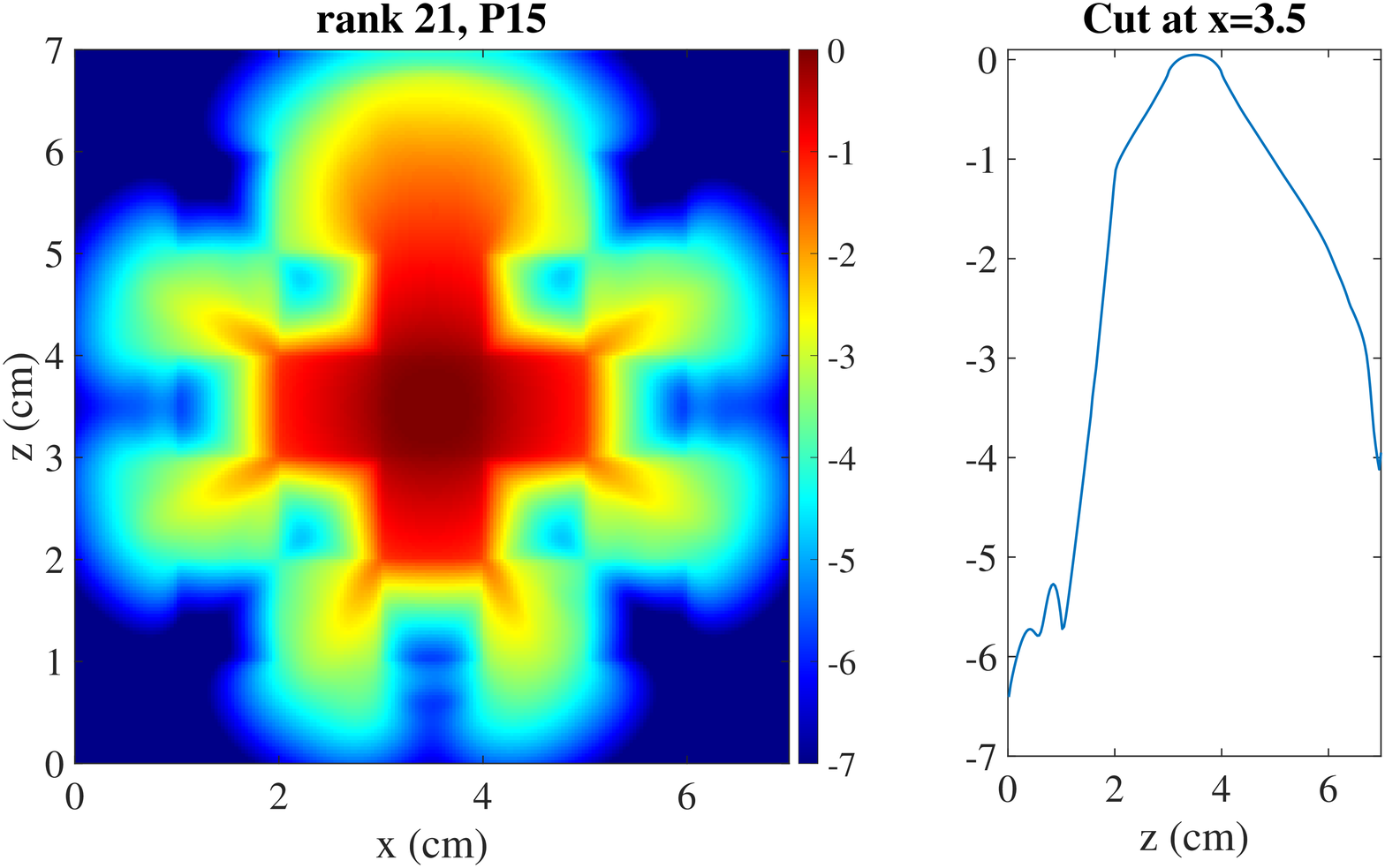}
	\end{subfigure}
	\begin{subfigure}{.5\textwidth}\centering
	\includegraphics[width=\columnwidth]{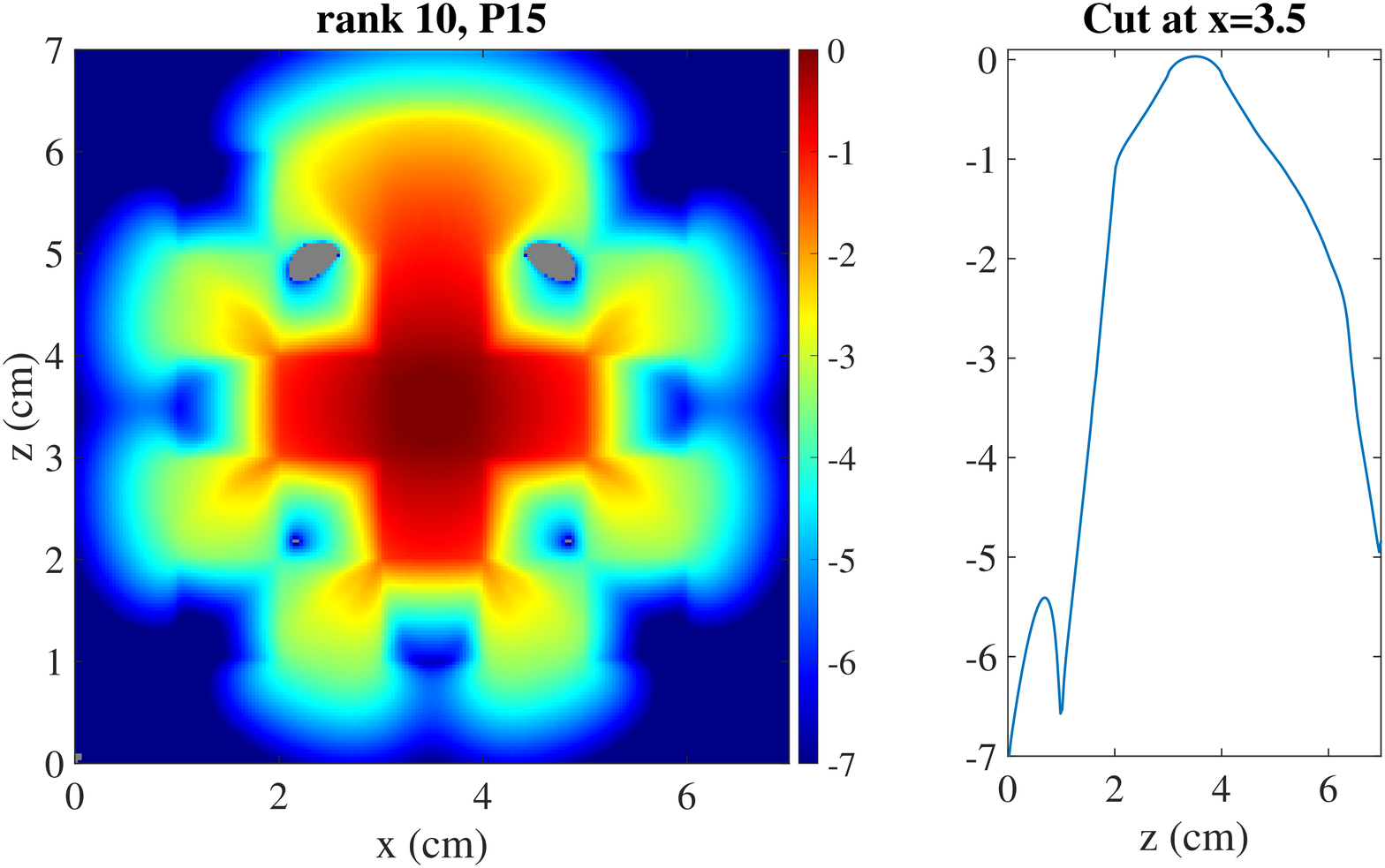}
	\end{subfigure}
	\hfill
	\begin{subfigure}{.5\textwidth}\centering
	\includegraphics[width=\columnwidth]{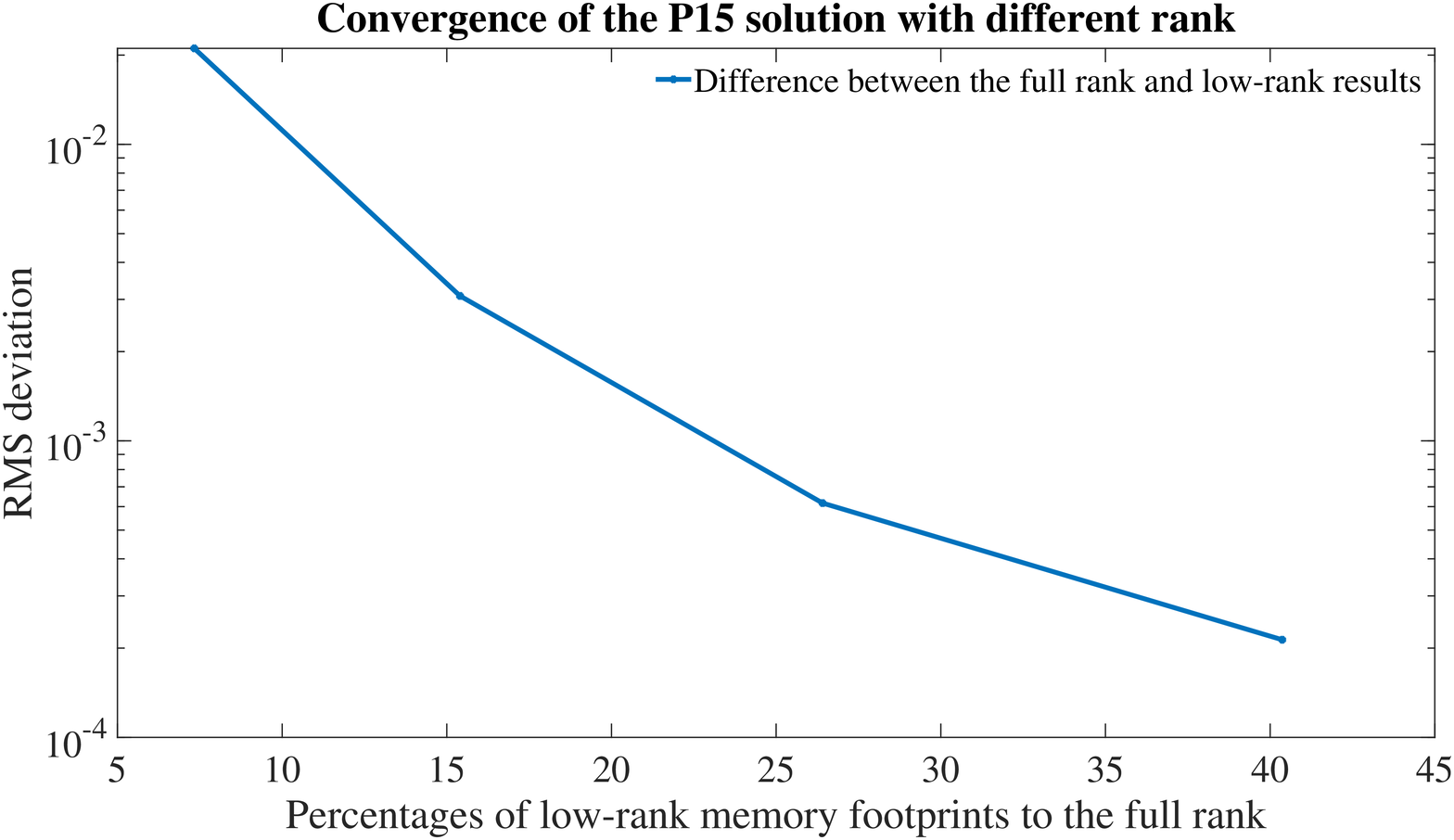}
	\end{subfigure}
	\caption{Solutions to Lattice problem at $t = 3.2 \,s$ with P$_{15}$ and different rank. The color scale is logarithmic.}
	\label{Lattice_results}
\end{figure}

\subsection{Lattice problem}
As described in Figure \ref{Lattice_layout}, the lattice problem is a $7 \times 7$ checkerboard with purely scattering zones $\sigmat = \sigmas = 1 \, $cm$^{-1}$, purely absorbing zones $\sigmat = 10\, \mathrm{cm}^{-1}, \sigmas = 0$, and an isotropic source at the center where $Q = 1$ \cite{Brunner2005}. In this problem, we use a spatial grid of size $210 \times 210$ for the computational domain $[0,7] \times [0,7]$.

Figure \ref{Lattice_results} shows that the solution with different rank give similar results except in the absorbing zones, where  solutions such as rank 55 and 36 contain more details than the rank 21 and 10 solutions. Additionally, low-rank solutions converge to the full rank as we increase the rank, which leads to more memory usage. As we can see that even with $26\%$ of the full rank memory the difference is still small. {Furthermore in this case the running time is $57\%$ shorter than the full rank solution as suggested in Table {\ref{Lattice_comparsion}}.}

\begin{table*}[hbt!]
\caption{\label{Lattice_comparsion}The quantitative comparison for the Lattice problem of memory footprints, running time, and errors with the same angular discretizations but different rank. }
\centering
\begin{adjustbox}{width=1.2\textwidth,center}
\begin{tabular}{|c|c|c|c|c|c|}
\hline
& \makecell{rank 136, $P_{15}$ \\ (full rank) } & \makecell{rank 10, $P_{15}$ \\ (low rank) } & \makecell{rank 21, $P_{15}$ \\ (low rank) } & \makecell{rank 36, $P_{15}$ \\ (low rank) } & \makecell{rank 55, $P_{15}$ \\ (low rank) }\\
\hline
\makecell{Theoretical \\ memory (MB)} & 95.96 & 7.08 & 14.87 & 25.50 & 38.98\\
\hline
\makecell{MATLAB \\memory (MB)} & 95.59 & 6.95 & 14.65 & 25.24 & 39.94\\
\hline
\makecell{Running time \\ per step (s)} & 3.19 & 0.45 & 0.84 & 1.36 & 2.03\\
\hline
\end{tabular}
\end{adjustbox}
\end{table*}

\begin{figure} 
	\centering
	\includegraphics[scale=0.5]{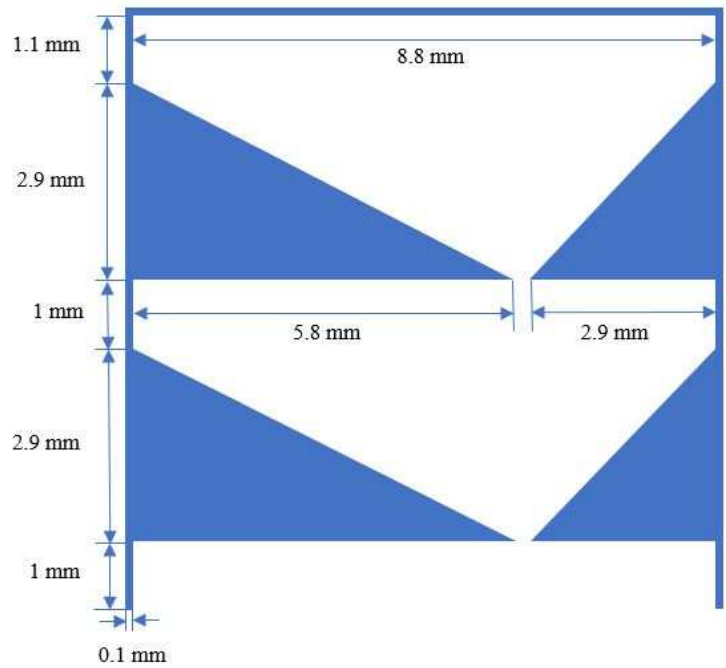}
	\caption{{The layout for the double chevron problem. The blue areas are walls and obstacles made up by dense materials. The blank is the purely-scattering dilute medium. The bottom has an isotropic incoming source boundary condition and the other three sides have vacuum boundary conditions.}}
	\label{Asym_layout}
\end{figure}

\begin{figure}[h!] 
	\begin{subfigure}{.5\textwidth}\centering
	\includegraphics[width=1.2\columnwidth]{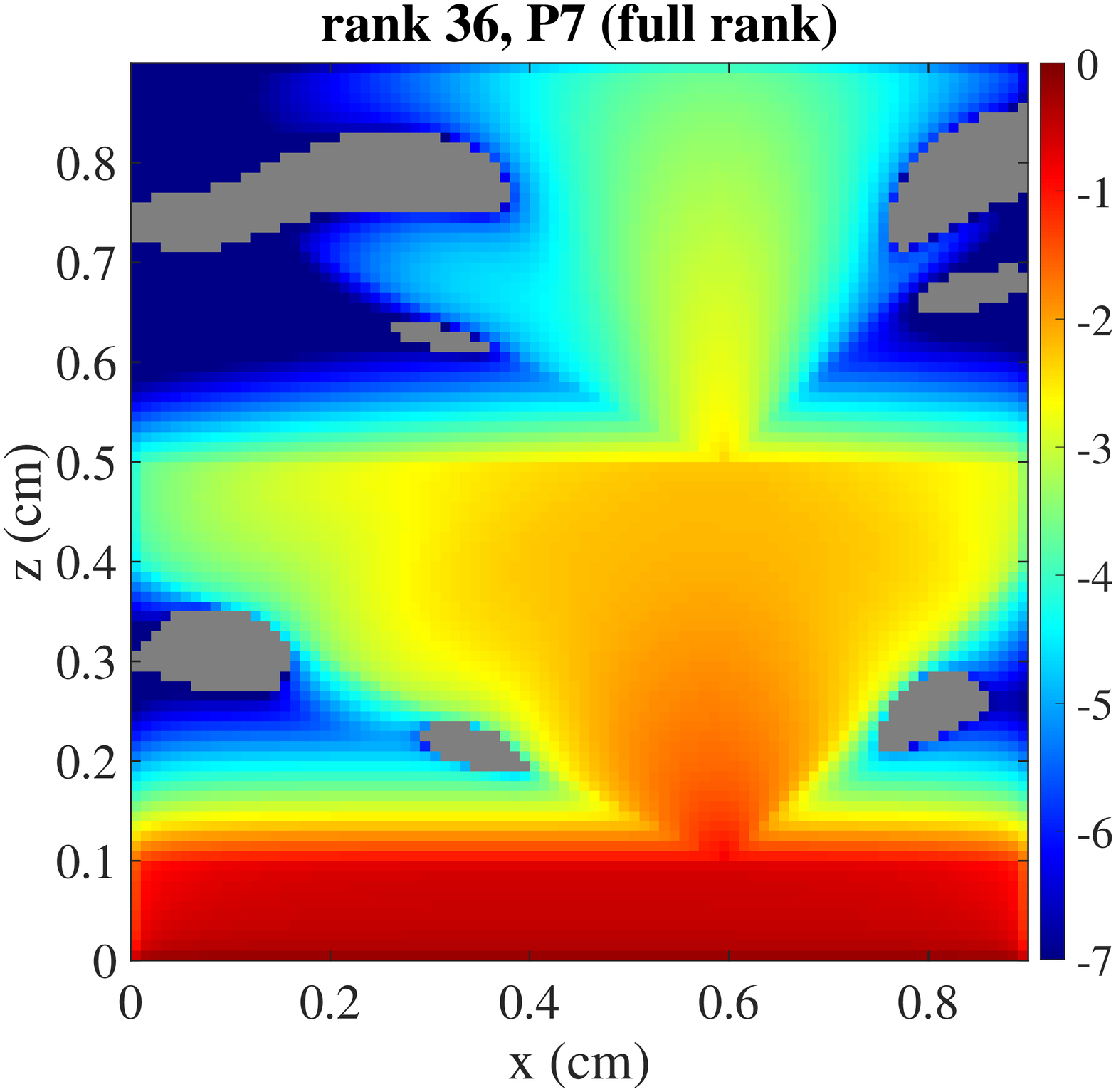}
	\end{subfigure}
	\hfill
	\begin{subfigure}{.5\textwidth}\centering
	\includegraphics[width=1.2\columnwidth]{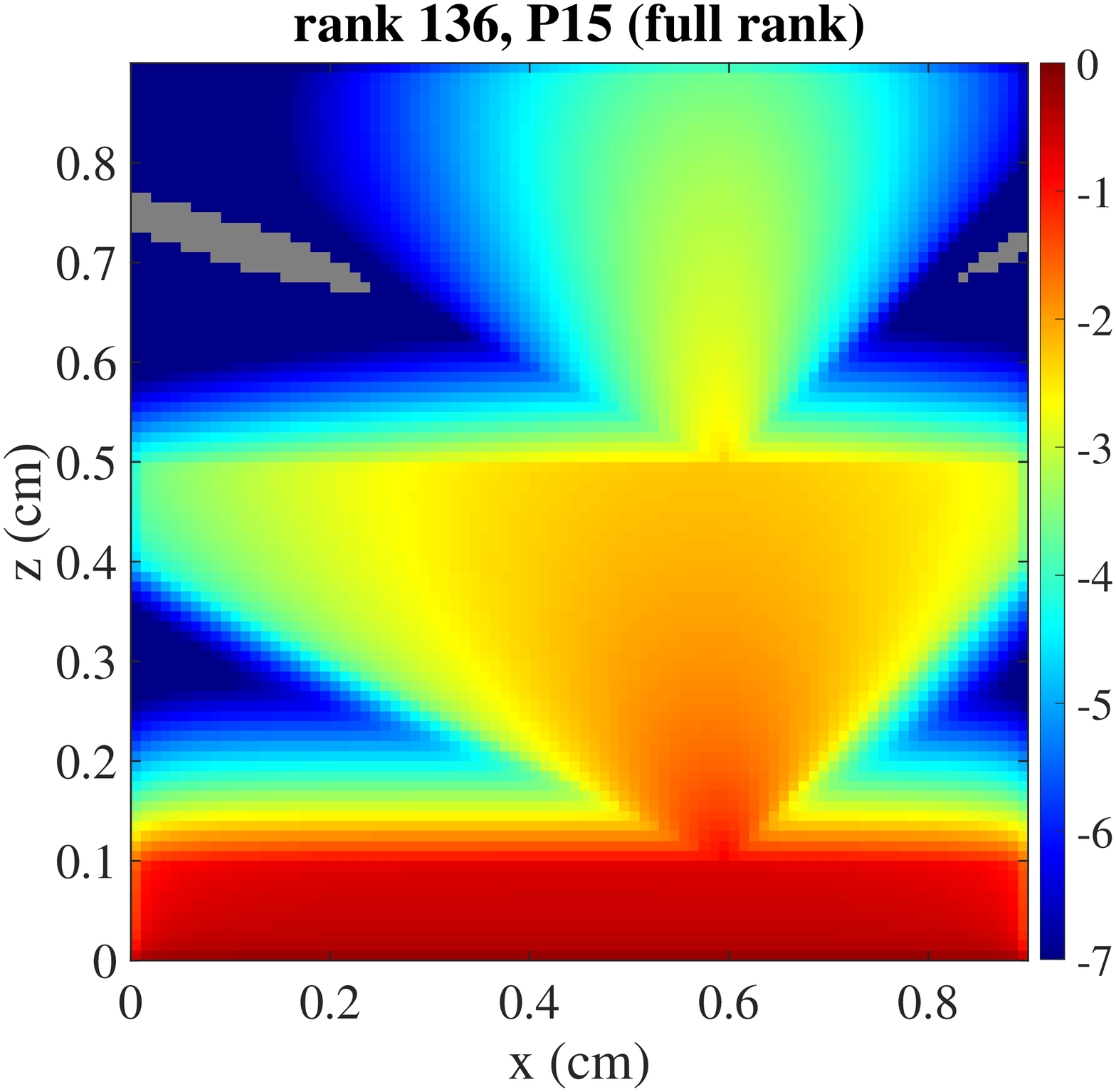}
	\end{subfigure}
	\begin{subfigure}{.5\textwidth}\centering
	\includegraphics[width=1.2\columnwidth]{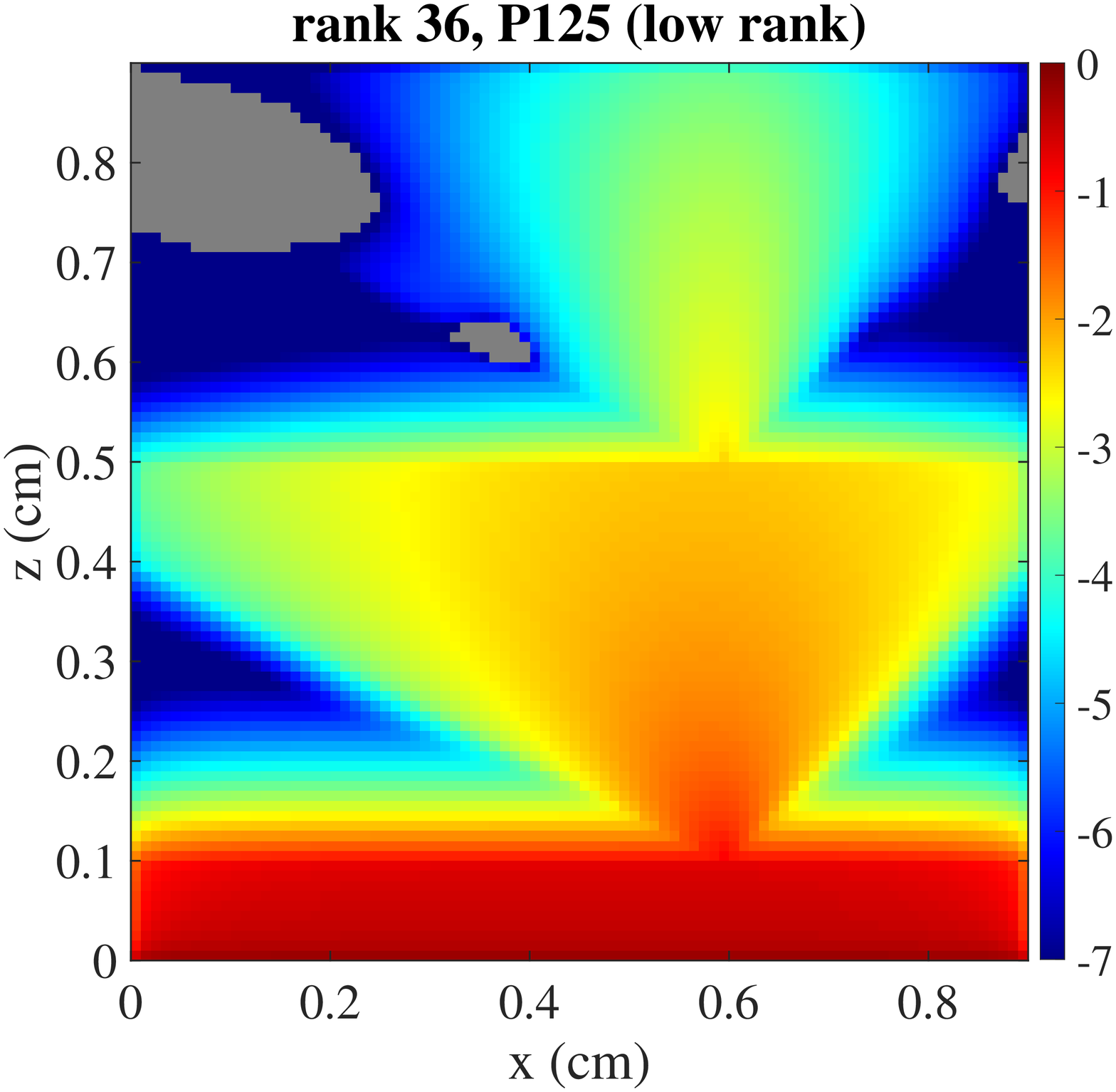}
	\end{subfigure}
	\hfill
	\begin{subfigure}{.5\textwidth}\centering
	\includegraphics[width=1.2\columnwidth]{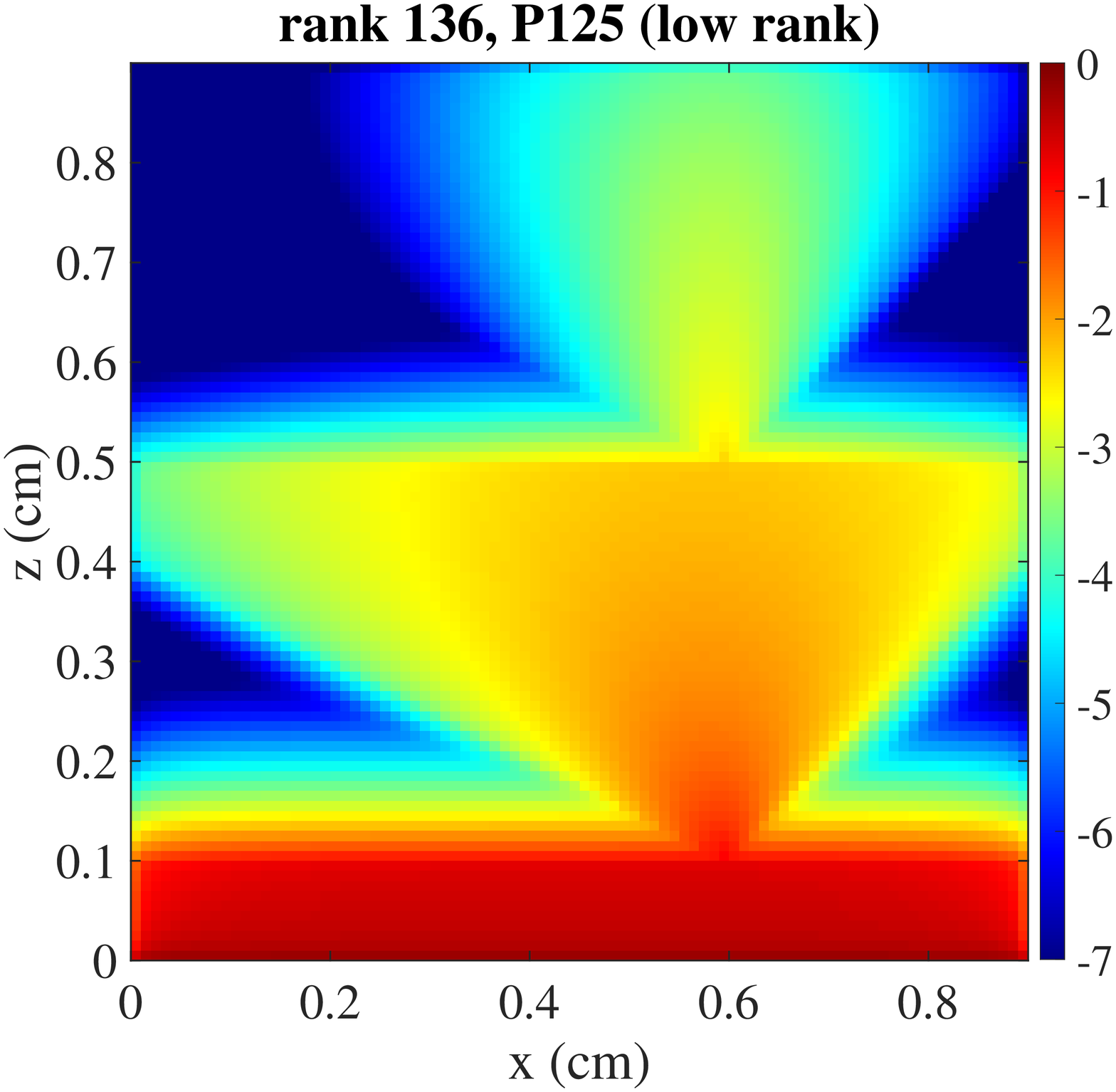}
	\end{subfigure}
	\begin{subfigure}{.5\textwidth}\centering
	\includegraphics[width=1.2\columnwidth]{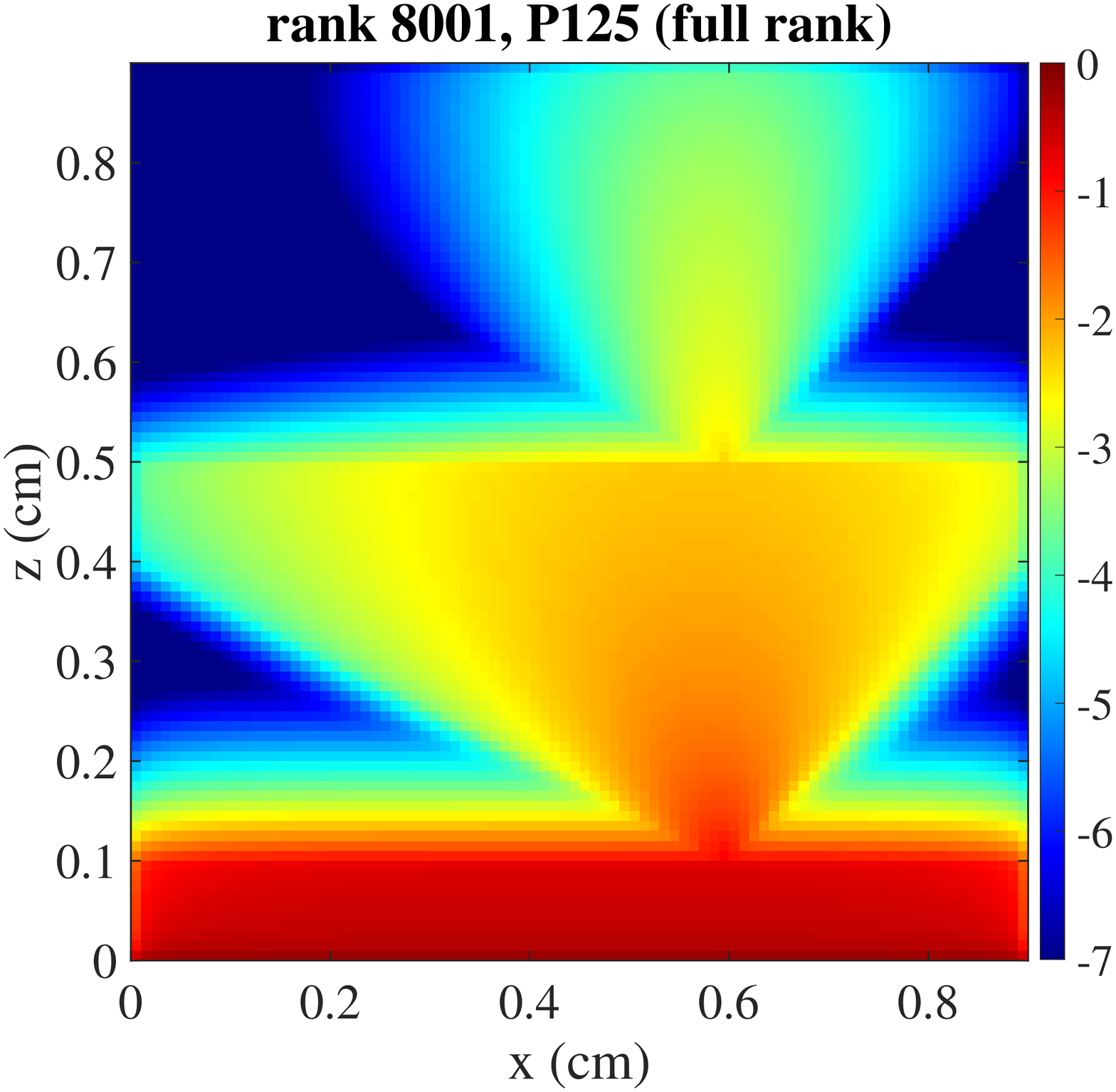}
	\end{subfigure}
	\hfill
	\begin{subfigure}{.5\textwidth}\centering
	\includegraphics[width=1.2\columnwidth]{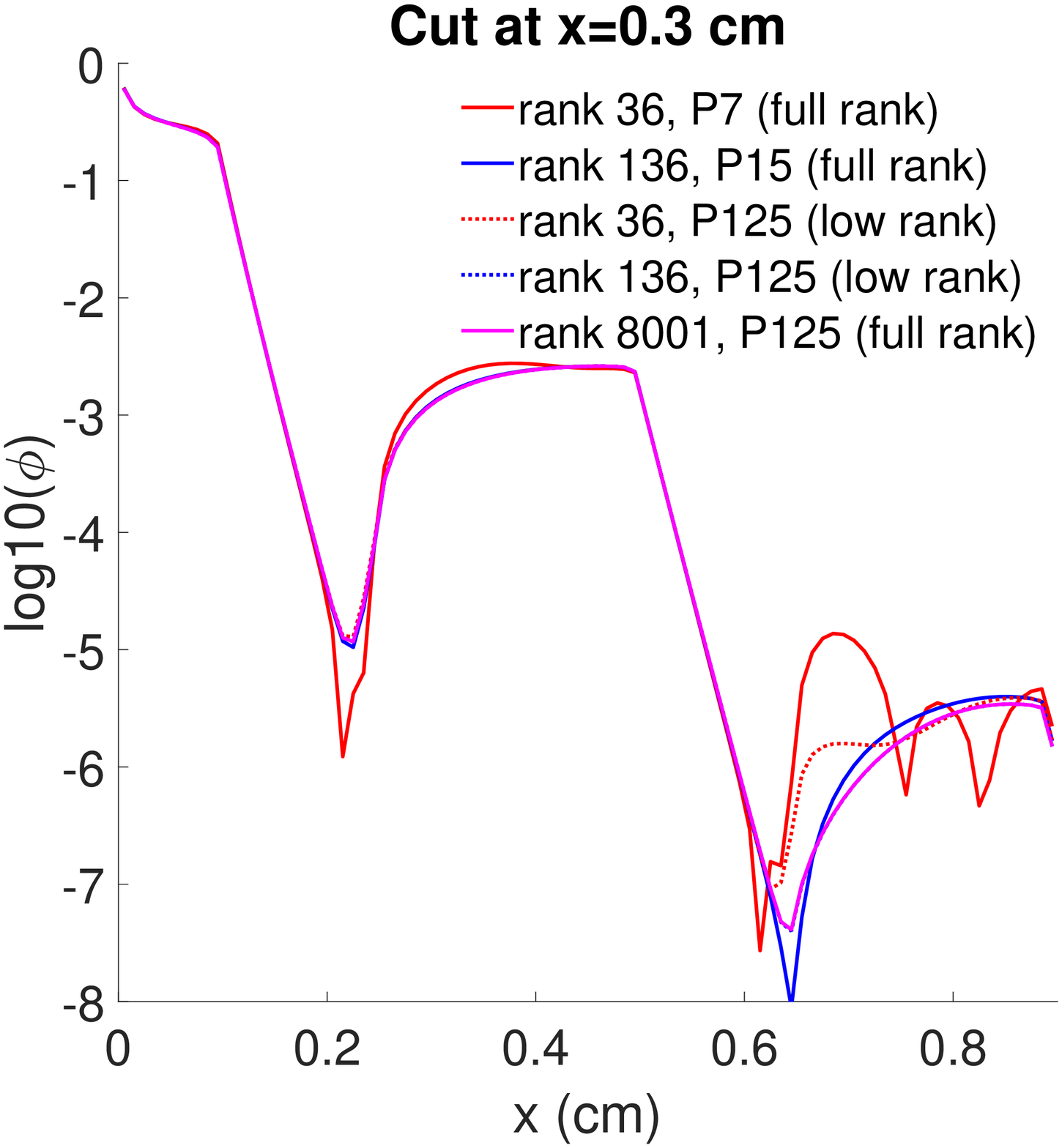}
	\end{subfigure}
	\caption{{Solutions to the double chevron problem at $t = 0.9 \,$s. The color scale is logarithmic and negative regions are shaded gray. Note the rank 136 $P_{125}$ solution is obscured by the full rank $P_{125}$ solution.}}
	\label{Asym_results}
\end{figure}

\begin{table*}[hbt!]
\caption{\label{DC_comparsion}The quantitative comparison for the double chevron problem of memory footprints, running time, and errors with the same rank but different angular discretizations.}
\centering
\begin{adjustbox}{width=1.4\textwidth,center}
\begin{tabular}{|c|c|c|c|c|c|}
\hline
& \makecell{rank 36, $P_7$ \\ (full rank) } & \makecell{rank 36, $P_{125}$ \\ (low rank) } & \makecell{rank 136, $P_{15}$ \\ (full rank) } & \makecell{rank 136, $P_{125}$ \\ (low rank) } & \makecell{rank 8001, $P_{125}$ \\ (full rank) }\\
\hline
\makecell{Theoretical \\ memory (MB)} & 4.67 & 9.29 & 17.63 & 35.33 & 1036.93\\
\hline
\makecell{MATLAB \\memory (MB)} & 4.50 & 9.09 & 17.48 & 34.76 & 1036.75\\
\hline
\makecell{Running time \\ per step (s)} & 0.26 & 1.16 & 0.64 & 2.31 & 34.48\\
\hline
\makecell{RMS deviation to\\ the full rank $P_{125}$ solution} & 3$\times 10^{-3}$ & 1.36$\times 10^{-5}$ & 6.74$\times 10^{-4}$ & 1.61$\times 10^{-8}$ & --\\
\hline
\end{tabular}
\end{adjustbox}
\end{table*}

\begin{figure} 
	\centering
	\includegraphics[scale=0.2]{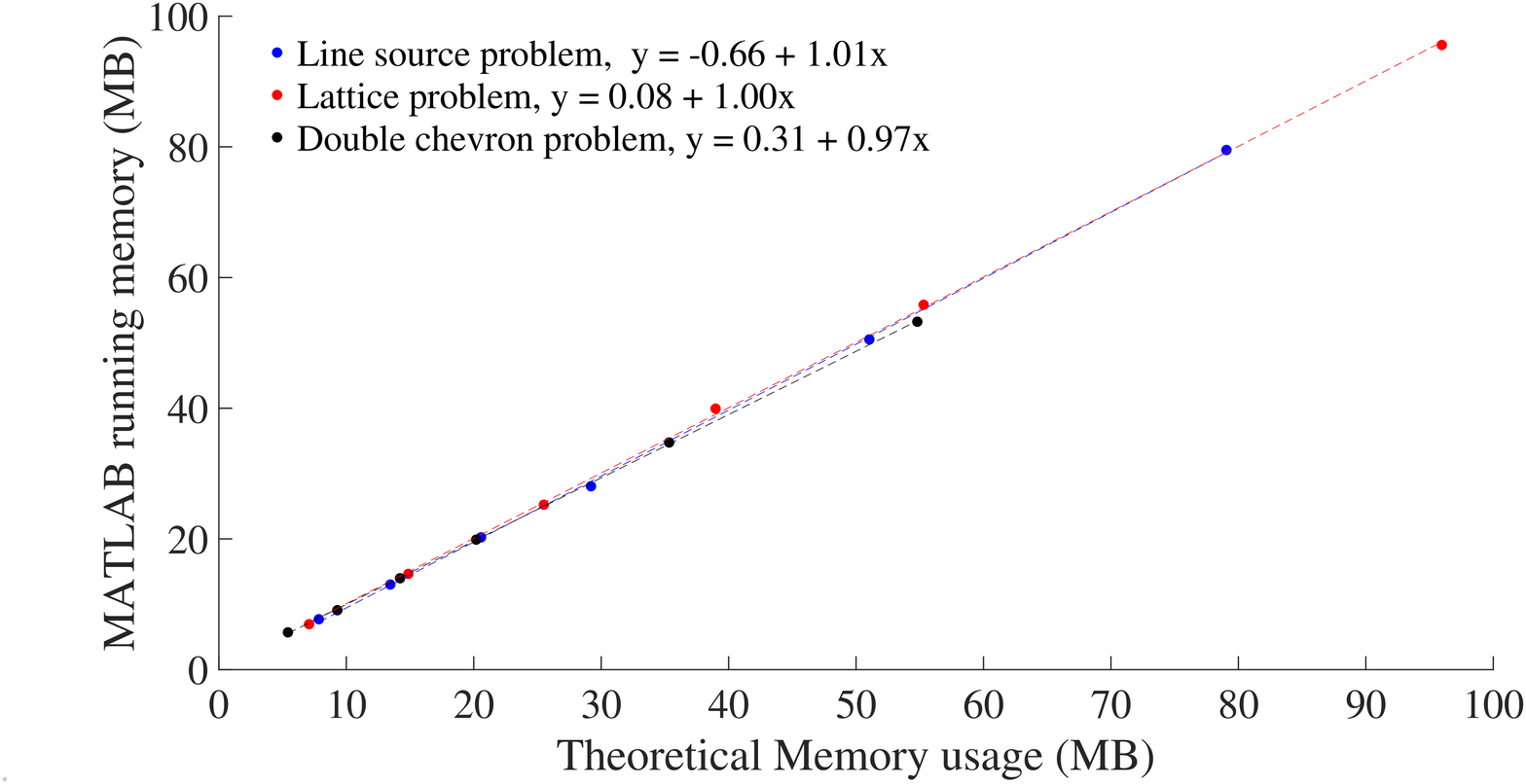}
	\caption{{The relationship between the memory calculated by Eqs. {\eqref{eq: 2D_lowrank_Memory}} and the running memory in MATLAB for the line source problem, the lattice problem and the double chevron problem. The angular discretization for the line source problem is P$_{39}$ with ranks varying from 21 to 210, the lattice problem is with P$_{15}$ and $r = 10, 21, 36, 55, 78, 136$ and the double chevron is with P$_{125}$ and the ranks are the same as the line source problem. The spatial and time discretizations are the same as previous simulations. The coefficients of determination ($R^2$) for the linear are approximately one, which indicates a strong linear relationship between the theoretical and actual memory for both problems.}}
	\label{memory_comparation}
\end{figure}

\subsection{Double Chevron problem} 

{In the last benchmark, we test our algorithm in an asymmetric geometry as shown in Figure $\ref{Asym_layout}$. The $9 \times 9$ spatial domain consists of purely scattering zones with $\sigmat = \sigmas = 0.01\, \mathrm{cm}^{-1}$, absorbing zones $\sigmat = 100\, \mathrm{cm}^{-1}, \sigmas = 0.1\, \mathrm{cm}^{-1}$, and an isotropic source at the bottom with $Q = 1$. In this problem, we use the spatial grid of size $90 \times 90$ for the computational domain $[0, 9\,\mathrm{mm}] \times [0, 9 \,\mathrm{mm}]$, and the simulation time $t = 0.9$ s. This problem is designed to show the benefits of our low rank method when $m \approx n$. Note that $m = 8100$ and $n = 8001$ with P$_{125}$. }

{It can be seen from Figure {\ref{Asym_results}} that the full rank P$_{125}$ solutionis identical in the scale of the figure to the rank 136 solution with P$_{125}$. Another observation we can make is that both the full rank P$_7$ and P$_{15}$ solutions are not able to capture the particle distribution behind the obstacles, leading to negative values, as indicated by the gray shading in the figure. Fortunately, these results can be improved if we refine the angular discretizations by applying the low-rank algorithm. Low-rank solutions with P$_{125}$ with the same rank as P$_7$ or P$_{15}$ are far better than the full-rank solutions. }

{The memory and running time for this problem is given in Table $\ref{DC_comparsion}$. We observe that the low-rank solution is far more accurate than the full rank solution with reasonable extra cost. For example, the error of the full rank P$_{15}$ solution can be reduced by a factor several thousand by employing the low rank method with $r=136$, which comes with a cost of only twice the memory and three times the computation time. Comparing to the full rank P$_{125}$ solution, this low rank solution only uses 3.4\% of the memory and 6.7\% of the computation time, but the overall error is small.}

\subsection{Theoretical versus Actual Memory Used}
{Figure {\ref{memory_comparation}} plots the actual memory occupied during the calculation with the theoretical values. Linear regression demonstrates that the memory used in our implementation is a linear function of the theoretical values. This verifies that the low-rank algorithm does reduce the memory requirement proportionally to the reduction in rank. Therefore, we are justified in using the theoretical memory usage estimates in our discussion and figures above.}  

\section{Conclusion and Future Work}
We have developed a practical algorithm to find the low-rank solution of the slab and planar geometry transport equation using explicit time integration. The method is based on projecting the equation to low-rank manifolds and numerically integrating with three steps. The numerical simulations show that on several test problems, a memory savings of about an order of magnitude can be achieved with the low-rank algorithm without sacrificing accuracy. This study establishes a projection-based framework for the direct model decomposition of the radiation transport problems.  

In future work will attempt to solve the conservation issues present in the low-rank method using a high-order low-order approach (such as quasi-diffusion), where the low-rank method is used only to create a closure. In this approach, the low-rank method is estimating a ratio of moments that should be insensitive to conservation issues. {Furthermore, there are several open research topics to explore, including asymptotic preservation, implicit discretization, the extension to energy-dependent problems, and other angular treatments, such as discrete ordinates. A recent preprint by Einkemmer, Hu, and Wang has presented a low-rank method based on an IMEX scheme that is asymptotic preserving {\cite{einkemmer2020asymptotic}}, and we believe this is a first of several new developments in low-rank algorithms for transport problems.} 

\nolinenumbers

\bibliography{mybibfile}

\begin{thebibliography}{10}
\expandafter\ifx\csname url\endcsname\relax
  \def\url#1{\texttt{#1}}\fi
\expandafter\ifx\csname urlprefix\endcsname\relax\def\urlprefix{URL }\fi
\expandafter\ifx\csname href\endcsname\relax
  \def\href#1#2{#2} \def\path#1{#1}\fi

\bibitem{Shalf2011}
J.~Shalf, S.~Dosanjh, J.~Morrison, {Exascale computing technology challenges},
  in: International Conference on High Performance Computing for Computational
  Science, 2010, pp. 1--25.

\bibitem{Pomraning}
G.~C. Pomraning, {The Equations of Radiation Hydrodynamics}, Dover
  Publications, 2005.

\bibitem{LewisMiller}
E.~Lewis, W.~Miller, Computational Methods of Neutron Transport, John Wiley and
  Sons, 1984.

\bibitem{Mcclarren2008}
R.~G. McClarren, T.~A. Brunner, J.~P. Holloway, {On solutions to the Pn
  equations for thermal radiative transfer}, Journal of Computational Physics
  227~(5) (2008) 2864 -- 2885.

\bibitem{Laboure:2016hm}
V.~M. Laboure, R.~G. McClarren, C.~D. Hauck, {Implicit filtered PN for
  high-energy density thermal radiation transport using discontinuous Galerkin
  finite elements}, Journal of Computational Physics 321 (2016) 624--643.

\bibitem{Dargaville:2019fa}
S.~Dargaville, A.~G. Buchan, R.~P. Smedley-Stevenson, P.~N. Smith, C.~C. Pain,
  {Angular adaptivity with spherical harmonics for Boltzmann transport},
  Journal of Computational Physics 397 (2019) 108846.

\bibitem{morel2000}
J.~E. Morel, {Diffusion-limit asymptotics of the transport equation, the
  P$_{1/3}$ equations, and two flux-limited diffusion theories}, Journal of
  Quantitative Spectroscopy and Radiative Transfer 65 (2000) 769--778.

\bibitem{Olson:2000vq}
G.~Olson, L.~Auer, M.~Hall, {Diffusion, P 1, and other approximate forms of
  radiation transport}, Journal of Quantitative Spectroscopy and Radiative
  Transfer 64~(6) (2000) 619--634.

\bibitem{McClarren:2011ga}
R.~G. McClarren, {Theoretical Aspects of the Simplified Pn Equations},
  Transport Theory and Statistical Physics 39~(2-4) (2011) 73--109.

\bibitem{Frank:2007ig}
M.~Frank, A.~Klar, E.~W. Larsen, S.~Yasuda, {Time-dependent simplified PN
  approximation to the equations of radiative transfer}, Journal of
  Computational Physics 226~(2) (2007) 2289--2305.

\bibitem{Modest:2012df}
M.~F. Modest, S.~Lei, {The Simplified Spherical Harmonics Method For Radiative
  Heat Transfer}, Journal of Physics: Conference Series 369 (2012) 012019.

\bibitem{anistratov1986solution}
D.~Anistratov, V.~Y. Gol’din, Solution of the multigroup transport equation
  by the quasidiffusion method, Preprint of the Keldysh Institute for Applied
  Mathematics 128 (1986) 19.

\bibitem{Fleck:1971vj}
J.~A. Fleck, J.~D. Cummings, {An implicit Monte Carlo scheme for calculating
  time and frequency dependent nonlinear radiation transport}, Journal of
  Computational Physics 8~(3) (1971) 313--342.

\bibitem{McClarren:2009bs}
R.~G. McClarren, T.~J. Urbatsch, {A modified implicit Monte Carlo method for
  time-dependent radiative transfer with adaptive material coupling}, Journal
  of Computational Physics 228~(16) (2009) 5669--5686.

\bibitem{Wollaber:2016hv}
A.~B. Wollaber, {Four Decades of Implicit Monte Carlo}, Journal of
  Computational and Theoretical Transport 45~(1--2) (2016) 1--70.

\bibitem{SmedleyStevenson:2015gt}
R.~P. Smedley-Stevenson, R.~G. McClarren, {Asymptotic diffusion limit of cell
  temperature discretisation schemes for thermal radiation transport}, Journal
  of Computational Physics 286 (2015) 214--235.

\bibitem{Koch2007}
O.~Koch, C.~Lubich, {Dynamical low-rank approximation}, SIAM Journal on Matrix
  Analysis and Applications 29~(2) (2007) 434--454.

\bibitem{Koch2010}
O.~Koch, C.~Lubich, {Dynamical Tensor Approximation}, SIAM Journal on Matrix
  Analysis and Applications 31~(5) (2010) 2360--2375.

\bibitem{Nonnenmacher2008}
A.~Nonnenmacher, C.~Lubich, {Dynamical low-rank approximation: applications and
  numerical experiments}, Mathematics and Computers in Simulation 79~(4) (2008)
  1346--1357.

\bibitem{Kloss2017}
B.~Kloss, I.~Burghardt, C.~Lubich, {Implementation of a novel
  projector-splitting integrator for the multi-configurational time-dependent
  Hartree approach}, Journal of Chemical Physics 146~(17).

\bibitem{KatharinaKormann2015}
{Katharina Kormann}, {A semi-Lagrangian Vlasov solver in tensor train format},
  SIAM Journal on Scientific Computing 37~(4) (2015) B613--B632.
\newblock \href {http://arxiv.org/abs/1408.7006} {\path{arXiv:1408.7006}},
  \href {http://dx.doi.org/10.1137/140971270} {\path{doi:10.1137/140971270}}.

\bibitem{Lubich2014}
C.~Lubich, I.~V. Oseledets, {A projector-splitting integrator for dynamical
  low-rank approximation}, BIT Numerical Mathematics 54~(1) (2014) 171--188.

\bibitem{Einkemmer2018}
L.~Einkemmer, C.~Lubich, {A low-rank projector-splitting integrator for the
  Vlasov--Poisson equation}, SIAM Journal on Scientific Computing 40~(5) (2018)
  B1330--B1360.

\bibitem{Einkemmer2019}
L.~Einkemmer, C.~Lubich, \href{https://doi.org/10.1137/18M1218686}{A
  quasi-conservative dynamical low-rank algorithm for the vlasov equation},
  SIAM Journal on Scientific Computing 41~(5) (2019) B1061--B1081.
\newblock \href {http://arxiv.org/abs/https://doi.org/10.1137/18M1218686}
  {\path{arXiv:https://doi.org/10.1137/18M1218686}}, \href
  {http://dx.doi.org/10.1137/18M1218686} {\path{doi:10.1137/18M1218686}}.
\newline\urlprefix\url{https://doi.org/10.1137/18M1218686}

\bibitem{Einkemmer2019_vlasovMaxwell}
L.~Einkemmer, A.~Ostermann, C.~Piazzola,
  \href{http://arxiv.org/abs/1902.00424}{{A low-rank projector-splitting
  integrator for the Vlasov--Maxwell equations with divergence correction}},
  Journal of Computational Physics 403 (2019) 109063.
\newblock \href {http://arxiv.org/abs/1902.00424} {\path{arXiv:1902.00424}},
  \href {http://dx.doi.org/10.1016/j.jcp.2019.109063}
  {\path{doi:10.1016/j.jcp.2019.109063}}.
\newline\urlprefix\url{http://arxiv.org/abs/1902.00424}

\bibitem{Einkemmer2019_weaklyflow}
L.~Einkemmer, {A low-rank algorithm for weakly compressible flow}, SIAM Journal
  on Scientific Computing 41~(5) (2019) A2795--A2814.

\bibitem{peng2019low}
Z.~Peng, R.~G. McClarren, M.~Frank, A low-rank method for time-dependent
  transport calculations, in: International Conference on Mathematics and
  Computational Methods applied to Nuclear Science and Engineering (M\&C 2019),
  2019.
\newblock \href {http://arxiv.org/abs/1906.09940} {\path{arXiv:1906.09940}}.

\bibitem{ding2019error}
Z.~Ding, L.~Einkemmer, Q.~Li, Error analysis of an asymptotic preserving
  dynamical low-rank integrator for the multi-scale radiative transfer equation
  (2019).
\newblock \href {http://arxiv.org/abs/1907.04247} {\path{arXiv:1907.04247}}.

\bibitem{Edelman1998}
A.~Edelman, T.~A. Arias, S.~T. Smith, {The Geometry of Algorithms with
  Orthogonality Constraints}, SIAM Journal on Matrix Analysis and Applications
  20~(2) (1998) 303--353.

\bibitem{Brown2005}
P.~N. Brown, D.~E. Shumaker, C.~S. Woodward, {Fully implicit solution of
  large-scale non-equilibrium radiation diffusion with high order time
  integration}, Journal of Computational Physics 204~(2) (2005) 760--783.

\bibitem{Brunner2005}
T.~A. Brunner, J.~P. Holloway, {Two-dimensional time dependent Riemann solvers
  for neutron transport}, Journal of Computational Physics 210~(1) (2005)
  386--399.

\bibitem{McClarren2010}
R.~G. McClarren, C.~D. Hauck, {Robust and accurate filtered spherical harmonics
  expansions for radiative transfer}, Journal of Computational Physics 229~(16)
  (2010) 5597--5614.

\bibitem{Radice2013}
D.~Radice, E.~Abdikamalov, L.~Rezzolla, C.~D. Ott, {A new spherical harmonics
  scheme for multi-dimensional radiation transport I. Static matter
  configurations}, Journal of Computational Physics 242~(SEPTEMBER 2012) (2013)
  648--669.

\bibitem{Raithby1999}
G.~D. Raithby, \href{https://doi.org/10.1080/104077999275631}{Evaluation of
  discretization errors in finite-volume radiant heat transfer predictions},
  Numerical Heat Transfer, Part B: Fundamentals 36~(3) (1999) 241--264.
\newblock \href {http://arxiv.org/abs/https://doi.org/10.1080/104077999275631}
  {\path{arXiv:https://doi.org/10.1080/104077999275631}}, \href
  {http://dx.doi.org/10.1080/104077999275631}
  {\path{doi:10.1080/104077999275631}}.
\newline\urlprefix\url{https://doi.org/10.1080/104077999275631}

\bibitem{Ganapol2008}
B.~Ganapol, Analytical Benchmarks for Nuclear Engineering Applications,
  Organisation for Economic Co-Operation and Development, 2008.

\bibitem{einkemmer2020asymptotic}
L.~Einkemmer, J.~Hu, Y.~Wang, An asymptotic-preserving dynamical low-rank
  method for the multi-scale multi-dimensional linear transport equation, arXiv
  preprint arXiv:2005.06571.

\end{thebibliography}

\end{document}